\documentclass{jfm}
\usepackage{graphicx}
\usepackage{epstopdf, epsfig}
\usepackage{xcolor} 
\usepackage{amsmath,amsbsy}

\shorttitle{Incompressible active phases at an interface}
\shortauthor{L. L. Jia, W. T. M. Irvine, and M. J. Shelley}

\title{Incompressible active phases at an interface. \\
I. Formulation and axisymmetric odd flows}

\author{Leroy L. Jia\aff{1}
  \corresp{\email{ljia@flatironinstitute.org}},
  William T. M. Irvine\aff{2}
 \and Michael J. Shelley\aff{1}\aff{3}}

\affiliation{\aff{1}Center for Computational Biology, Flatiron Institute, New York, NY 10010, USA
\aff{2}James Franck Institute, Enrico Fermi Institute, and Department of Physics, University of Chicago, Chicago, IL 60637, USA
\aff{3} Courant Institute of Mathematical Sciences, New York University, New York, NY 10012, USA}

\begin{document}

\maketitle

\begin{abstract}
Inspired by the recent realization of a 2D chiral fluid as an active monolayer droplet moving atop a 3D Stokesian fluid, we formulate mathematically its free-boundary dynamics. The surface droplet is described as a general 2D linear, incompressible, and isotropic fluid, having a viscous shear stress, an active chiral driving stress, and a Hall stress allowed by the lack of time-reversal symmetry. The droplet interacts with itself through its driven internal mechanics and by driving flows in the underlying 3D Stokes phase. We pose the dynamics as the solution to a singular integral-differential equation, over the droplet surface, using the mapping from surface stress to surface velocity for the 3D Stokes equations. Specializing to the case of axisymmetric droplets, exact representations for the chiral surface flow are given in terms of solutions to a singular integral equation, solved using both analytical and numerical techniques. For a disc-shaped monolayer, we additionally employ a semi-analytical solution that hinges on an orthogonal basis of Bessel functions and allows for efficient computation of the monolayer velocity field, which ranges from a nearly solid-body rotation to a unidirectional edge current depending on the subphase depth and the Saffman-Delbr\"uck length. Except in the near-wall limit, these solutions have divergent surface shear stresses at droplet boundaries, a signature of systems with codimension one domains embedded in a three-dimensional medium. We further investigate the effect of a Hall viscosity, which couples radial and transverse surface velocity components, on the dynamics of a closing cavity.

\end{abstract}


\section{Introduction}
\label{sec:intro}

In this work, we develop a mathematical description of the free-boundary dynamics of a 2D incompressible droplet moving atop a bulk Stokes fluid. Following the approach of \cite{Soni_etal2019}, this incompressible surface phase can be either active or passive, and is described by the most general linear isotropic fluid model. This model allows a viscous shear stress, an antisymmetric "chiral" stress reflecting the driven rotation of the fluid constituents, and an "odd" Hall stress allowed by the consequent loss of time-reversal symmetry at the microscopic level. Given its generality, this model touches upon both classical and emerging areas of fluid dynamics and applied mathematics, including Langmuir films, mixed dimension boundary value problems, Euler and quasigeostrophic vortex systems, and active matter systems. We briefly describe connections with these areas before stating our main results.

The interaction of rotating elements in a fluid is a foundational topic in fluid dynamics, going back to the explication of 2D point vortices of Euler equations interacting through the Biot-Savart law~(\cite{Saffman1995}). A 2D patch of constant vorticity interacts with itself similarly and its dynamics can be reduced to a free-boundary problem~(\cite{Pullin1992}). The surface quasi-geostrophic equations (SQG) of atmospheric physics have their own singular and free-boundary analogues (\cite{HPGS1995,RF2004}). Rotational interaction problems also arise, in the guise of so-called active matter, in the zero Reynolds limit of the Stokes equations where solid particles are driven to rotate by an external field, or through internal actuation. Ensembles of such particles will interact through their induced fluid flows, steric interactions, and possibly other fields such as magnetic. These systems can show activity-induced phase separation~(\cite{Yeo_etal2015}), crystallization and hyperuniformity~(\cite{Petroff_etal2015,Oppenheimer_etal2019,Oppenheimer_etal2022}), odd surface waves and edge currents~(\cite{Soni_etal2019}), complex interactions of vortical structures~(\cite{BililignEtAl2021}), and forms of active turbulence~(\cite{Kokot_etal2017}).

When such many-particle systems are modeled as continuous fluidic materials, novel internal stresses can arise. Firstly, the driven rotation of the fluid's constituents gives rise to an anti-symmetric driving stress. Consequent to the microscopic driving of rotation these out-of-equilibrium fluids do not obey time-reversal symmetry and so can possess a so-called odd or Hall stress which, in its simplest case, is linear in rates-of-strain and couples longitudinal and transverse flow components. Examples of such systems include quantum Hall fluids (\cite{Avron_etal1995}), vortex fluids (\cite{WiegmannAbanov2014}), and electron fluids in graphene (\cite{Berdyugin_etal2019}). Fluids with an odd viscosity can exhibit rheological properties and exotic flow phenomena markedly different from their Newtonian counterparts such as unidirectional edge currents or topological waves (\cite{Souslov_etal2019,Soni_etal2019}).

Many of the examples above are of rotor assemblies sitting on a 2D fluid interface either embedded within, or sitting atop, a viscous fluid bulk. Other active matter systems, particularly active nematics formed of microtubule bundles and molecular motors, have been studied in this geometry both experimentally (see, e.g. \cite{SanchezEtAl2012}) and theoretically (see, e.g. \cite{GBGBS2015a}). These systems have generally been modelled as a 2D incompressible active material covering the entire surface, and the bulk as an incompressible Stokes fluid driven by the surface shear-stress. It has been shown that the bulk flows can profoundly modify the active surface dynamics, for example by introducing new length-scales of system instability at the onset of active nematic turbulence~(\cite{GBJS2017,MartinezEtAl2019}). A complementary line of study concerns the turbulent statistics and intermittency of flows within a flat surface that overlay a turbulent and incompressible 3D bulk~(\cite{GoldburgEtAl2001,CG2004}). Some part of the complexity of surface flows in this case derives from the 2D compressibility of the 2D surface velocity in their reflection of 3D inertial turbulence. 

\cite{Soni_etal2019} studied theoretically the dynamics of active chiral surface droplets where the contribution of the underlying bulk fluid was modeled as a simple local drag term, as is appropriate for the dynamics of large droplets near a solid substrate. This yields a homogeneous Brinkman equation, with activity-driven boundary conditions, for the droplet's in-plane velocity field. The kinematic boundary condition then evolves the droplet domain. Interactions through the bulk fluid, much less with other droplets, are completely screened in this near wall limit. Here we allow full coupling between the surface phase and the bulk fluid subphase, allowing the droplet to interact with itself both through its internal stresses, and through induced 3D fluid motions. In our formulation we use the Neumann to Dirichlet map for the 3D Stokes equations in a finite-depth layer or half-space (\cite{MasoudShelley2014}) to express the surface velocity as a surface convolution of a singular kernel with the surface stress, with that shear stress produced by the surface phase. This relation is quite general and here gives rise to a difficult and novel free-boundary problem. 

As first exploratory problems, we restrict our study here to axisymmetric solutions. To study rotational flows we determine the activity-driven flows within circular droplets and in the bulk. To study moving interfaces, we also study domains with holes (annuli) and study the course of hole closure and how it is affected by system parameters. Our analyses and computations thereof show that the bulk surface shear stresses diverge as an inverse square-root at the droplet boundary. Nonetheless, despite the divergence, the in-plane velocities remain bounded and continuous. For the disk, this divergence is associated with the rotational drive. Singular flows arise in other rotational systems, such as for an infinitely thin solid disk rotating in a Stokes fluid~(\cite{Jeffery1915}) whose edge shear stress diverges similarly, and for the SQG vortex patch problem~(\cite{RF2004}), which exhibits logarithmic divergences in tangential surface velocity. An identical logarithmic divergence is found, in a continuum limit, for planar assemblies of rotating particles rotating in a Stokes fluid~(\cite{YanEtAl2020}).

This work also adds to classic work in applied mathematics on the solution of three-dimensional mixed boundary value problems in potential theory, which arise when solving elliptic problems on two or more different domains, each of which has a different boundary condition that must be satisfied. Such problems, in axisymmetric settings, are frequently converted into multiple integral equations, and powerful methods have been developed to extract their near-analytical solutions; see~\cite{Sneddon1966} for an overview. These techniques are not readily applicable as our inhomogeneous forcing does not come from a prescribed stress or velocity field on any domain. Since the boundary includes a two-dimensional fluid monolayer, there is an additional two-dimensional elliptic problem with its own higher codimension boundary conditions coupled to the base three dimensional one. Other authors have more recently examined related problems in similar geometries but included simplifying assumptions such as the complete absence of a vertical flow component (\cite{StoneMcConnell1995}) or a vanishing monolayer viscosity that reduces the order of the monolayer equation (\cite{Alexander_etal2006}). Here, we preserve full generality and formulate a dual integral equation -- in addition to the Green's function formulation -- to obtain a near-analytical solution for a circular droplet, and demonstrate the formulation as triple integral equations for flow in an annulus. 

We begin by giving the mathematical formulation and governing equations in \textsection\ref{sec:mathmodel}. The solution to the general mathematical problem is stated in terms of a Green's function. We then proceed to specialize the formulation to the axisymmetric case in  \textsection\ref{sec:physsystem}. In \textsection \ref{sec:disc} and \textsection \ref{sec:annulus}, we demonstrate the solutions to the discal and annular geometries. An appendix reviews the experimental system and parameter values, lists the nondimensional groups associated with the parameters, and considers the solution in the infinite strip geometry.

\section{A mathematical model}
\label{sec:mathmodel}

We consider a surface phase domain $\mathcal{D}$ on the upper surface, $z=0$, of a layer of passive 3D Stokes fluid (subphase) of depth $H$ and infinite extent in the $x$ and $y$ directions. Gravitational forces and curvature of the interface are ignored, and we assume that the vertical velocity vanishes at the surface $z=0$. At $z=-H$, the 3D fluid subphase is in contact with a wall where it satisfies a no-slip condition. We assume that the 3D velocity field $\boldsymbol{u}$ and pressure field $p$ of the subphase satisfy the incompressible Stokes equations
\begin{equation}
-\bnabla_{3D} p + \mu \bnabla_{3D}^2\boldsymbol{u} = 0 
~~\text{  and  }~~  
\bnabla_{3D} \bcdot \boldsymbol{u}= 0,
\label{3DStokes}
\end{equation}
where $\mu$ is the viscosity of the subphase and $\bnabla_{3D}$ is the three-dimensional gradient operator $(\p/\p x, \p /\p y, \p/\p z)$.

Let $\boldsymbol{U}$ be the 2D fluid velocity field in the $z=0$ plane. The surface and bulk velocities are related by continuity: $\boldsymbol{u}(x,y,z=0) = (\boldsymbol{U}(x,y),0)$, a notation that captures the condition that the surface remains flat. Following \cite{Soni_etal2019}, we take the surface phase in $\mathcal{D}$ to be described by a general incompressible and isotropic 2D fluid with linear viscous and Hall stresses, and driven by an anti-symmetric stress. And so, firstly, we have
\begin{equation}
\bnabla\bcdot\boldsymbol{U} = 0, \quad \boldsymbol{x}\in \mathcal{D},
\label{divUeqn}
\end{equation}
where $\bnabla=(\p/\p x , \p/\p y)$ is the two-dimensional gradient on the surface. Secondly, the stress tensor, $\boldsymbol{\sigma}$, of the 2D active surface phase takes the form from \cite{Soni_etal2019}:
\begin{equation}
\boldsymbol{\sigma} = - P\mathsfbi{I} + \eta (\bnabla \boldsymbol{U} + \bnabla \boldsymbol{U}^T) + \eta_R(2\Omega - \omega) \mathsfbi{J} + \eta_O(\bnabla^\perp \boldsymbol{U} + \bnabla \boldsymbol{U}^\perp)
\label{sigmaeqn}
\end{equation}
where $P$ is the planar pressure enforcing that $\bnabla\bcdot\boldsymbol{U} = 0$, $\omega = \boldsymbol{e}_z \cdot (\bnabla_{3D}\times \boldsymbol{U} )$ is the scalar vorticity, and  $\Omega$ is the rotation frequency of the external magnetic field, taken to be spatially uniform and time-independent. The tensors $\mathsfbi{I}$ and $\mathsfbi{J}$ are the two-dimensional identity and anti-symmetric Levi-Civita tensors, respectively, while the operator $\perp$ maps a 2D vector to its rotation by $\upi/2$, i.e. $(v_1,v_2)^\perp=(-v_2,v_1)$. There are three different viscous moduli in our model: $\eta$ is a standard shear viscosity that arises from overcoming the magnetic attraction between nearby dipoles in relative motion, $\eta_R$ is the rotational viscosity that models friction between neighboring rotating particles, and $\eta_O$ is the odd viscosity (also known as the Hall viscosity) that gives rise to viscous forces acting transversely to a velocity gradient. As first noted in~\cite{Avron1998}, the odd viscous stress is a non-dissipative term that is only permissible in two-dimensional fluids that do not obey local  time-reversal symmetry. Its presence is intimately tied to the anti-symmetric driving stress.

The transverse motion of the surface phase droplet generates a shear stress on the bulk fluid below, and so we have for $\boldsymbol{u} = (u_1,u_2,u_3)$,
\begin{equation}
\bnabla\bcdot \boldsymbol{\sigma}=\left.\mu\frac{\partial(u_1,u_2)}{\partial z} \right|_{z=0}=\boldsymbol{f}, \quad \boldsymbol{x}\in \mathcal{D},
\label{momentumeqn1}
\end{equation}
which can be interpreted as a boundary condition on the subphase. Outside of the surface phase domain $\mathcal{D}$ we have a simple stress-free boundary condition on the subphase, or
\begin{equation}
\boldsymbol{f} = \boldsymbol{0}, \quad \boldsymbol{x}\notin \mathcal{D}.
\label{nostresseqn}
\end{equation}
Hence, Eqns~(\ref{momentumeqn1}) and (\ref{nostresseqn}) can be combined using a characteristic function $\chi$, 
\begin{equation}
\left.\mu\frac{\partial (u_1,u_2)}{\partial z} \right|_{z=0} = \chi({\mathcal{D}})(\bnabla\bcdot \boldsymbol{\sigma})~.
\label{momentumeqnchi}
\end{equation}
Here it has been assumed that the normal stress of the bulk phase at $z=0$ is whatever it needs to be to maintain surface flatness. This could be achieved, for example, by having a high surface tension there.

The expression for $\bnabla\bcdot \boldsymbol{\sigma}$ has a remarkably simple form. Using the notation of the skew gradient, we note that the scalar vorticity can be written as $\omega  = \nabla^\perp \bcdot \boldsymbol{U}$. It then follows, using Eqn.~(\ref{divUeqn}), that $\bnabla\bcdot (\omega \mathsfbi{J}) = -\nabla^\perp(\nabla^\perp\bcdot \boldsymbol{U})=-\Delta \boldsymbol{U}$. Similar manipulations yield the identities $\bnabla\bcdot(\bnabla^\perp \boldsymbol{U})=\boldsymbol{0}$ and $\bnabla\bcdot (\bnabla \boldsymbol{U}^\perp) = -\bnabla \omega$ for the divergence of the odd viscous tensor, which leads to the important consequence that the effect of the odd viscous stress in the bulk is simply to generate a ``pressure field'' proportional to vorticity. In fact, it is convenient to define $\bar\eta = \eta + \eta_R$ and $P \to P + \eta_O\omega$ so that we may write
\begin{equation}
\bnabla\bcdot \boldsymbol{\sigma}=-\nabla P + \bar\eta \bnabla^2\boldsymbol{U},\quad \boldsymbol{x}\in \mathcal{D}.
\label{divsigmaeqn}
\end{equation}
That is, the active phase is described by a 2D Stokes equation with the viscosity and pressure redefined. The fact that Eqn.~(\ref{divsigmaeqn}) does not depend explicity on $\Omega$ or $\eta_O$ implies that the drive and Hall stress can appear only through boundary conditions. 
Thus, we have 
\begin{equation}
   \chi(\mathcal{D})(-\nabla P + \bar\eta \bnabla^2\boldsymbol{U}) =\left.\mu\frac{\partial(u_1,u_2)}{\partial z} \right|_{z=0},
   \label{momentumeqn}
\end{equation}
coupled to the Stokes equations (\ref{3DStokes}) for the bulk phase.

The surface Stokes equation requires additional, transverse boundary conditions. We impose a stress balance condition on the surface phase boundary $\partial \mathcal{D}$,
\begin{equation}
\boldsymbol{\sigma} \bcdot \hat{\boldsymbol{n}} = \gamma \kappa \hat{\boldsymbol{n}},\quad \boldsymbol{x} \in \partial \mathcal{D}
\label{stressbcedge}
\end{equation}
where $\gamma$ is the surface tension, $\kappa$ is the local curvature of $\partial\mathcal{D}$, and $\hat{\boldsymbol{n}}$ is its inward facing normal vector (in the $z=0$ plane).
In the local Frenet frame of $\partial \mathcal{D}$ where $\boldsymbol{U} = T\hat{\boldsymbol{t}} + N\hat{\boldsymbol{n}}$, we find that $(\bnabla \boldsymbol{U} +\bnabla \boldsymbol{U}^T)\bcdot \hat{\boldsymbol{n}} = 2 \boldsymbol{U}_s^\perp-\omega \hat{\boldsymbol{t}}$, $\mathsfbi{J}\cdot \hat{\boldsymbol{n}} = \hat{\boldsymbol{t}}$, and $(\bnabla^\perp \boldsymbol{U} + \bnabla \boldsymbol{U}^\perp)\bcdot \hat{\boldsymbol{n}} = 2\boldsymbol{U}_s - \omega \hat{\boldsymbol{n}}$ so that 
{Eqn.~(\ref{stressbcedge}) can be written as}
\begin{equation}
    \left.-P + 2\eta (T_s - \kappa N) + 2\eta_O(N_s + \kappa T) \right|_{\partial \mathcal{D}} = \left. \gamma \kappa \right|_{\partial \mathcal{D}}
    \label{frenetbcrad}
\end{equation}
\begin{equation}
   \left. -\bar\eta \omega - 2\eta (N_s +\kappa T) + 2\eta_O (T_s-\kappa N) \right|_{\partial \mathcal{D}}= \left. - 2\eta_R \Omega \right|_{\partial\mathcal{D}}
   \label{frenetbctan}
\end{equation}
where the subscript $s$ denotes differentiation in the direction of $\hat{\boldsymbol{t}}$. This form of the boundary condition makes it clear that the role of the odd viscosity is to ``complement'' the shear viscosity by producing a tangential boundary stress that depends on the local normal velocity and vice versa, thus coupling the flows in the two directions. 
We remark that the $\Omega$ term in Eqn.~(\ref{frenetbctan}) provides the inhomogeneous forcing through which a nontrivial solution arises.

The dynamics of the active surface phase is a free boundary problem for $\partial\mathcal{D}$. If its velocity is $\boldsymbol{V}$ then we evolve $\partial\mathcal{D}$ through the kinematic boundary condition $\boldsymbol{V}=\boldsymbol{U}|_{\boldsymbol{x}\in\partial\mathcal{D}}$ (implicitly assuming continuity and boundedness of $\boldsymbol{U}$). Equations (\ref{3DStokes}-\ref{sigmaeqn}), (\ref{momentumeqnchi}), and (\ref{stressbcedge}), together with the conditions $\boldsymbol{\hat z}\cdot\boldsymbol{u}|_{z=0}=0$ and continuity of subphase and surface phase horizontal velocities at $z=0$, yield a complex, but complete, formulation for the determination of $\boldsymbol{U}$.

{\bf Some length and time scales:}
The many parameters of the model give rise to a number of relevant length and time scales, a detailed analysis of which is provided in Appendix B. Here we mention of few. One important length scale, independent of geometry and activity, is the Saffman-Delbr\"uck length $\ell_{SD} = \eta /\mu$ (\cite{SaffmanDelbruck1975}). On length scales smaller than $ \ell_{SD}$, momentum travels primarily in the plane of the surface phase, while for length scales larger than $\ell_{SD}$, momentum travels through the subphase as well.
For problems in which the characteristic size is variable, such as the edge tension-driven closure of a cavity punctured in the monolayer, the dynamics may take place in both regimes (\cite{JiaShelley2022}). A related length scale for surface phase droplets very close to the bottom wall is the penetration depth ${\bar\delta}=(H{\bar\eta}/\mu)^{1/2}$. This is the length scale of edge currents driven by the rotational drive (\cite{Soni_etal2019}). One obvious time-scale is that of the rotational drive $\tau_1=\Omega^{-1}$, while another is the relaxational time-scale $\tau_2=\eta R/\gamma$ driven by surface tension. Both arise in the solutions constructed herein, even though the time-scale for rotation is about one-hundredth of that for relaxation (that is, for the experiments of \cite{Soni_etal2019}).

\subsection{A Green's function formulation for the free-boundary problem.}

The infinite horizontal extent of the domain makes the problem amenable to classical Fourier transform methods. \cite{MasoudShelley2014} showed that in this geometry, 2D Fourier transforming Eqn.~(\ref{momentumeqn1}) in $x$ and $y$ and applying  boundary conditions yields the relation
\begin{equation}
\hat{\boldsymbol{f}} = \mu k\left[A(kH) \mathsfbi{I} + B(kH)\hat{\boldsymbol{k}} \hat{\boldsymbol{k}} \right]\hat{\boldsymbol{U}}
\label{fteqn}
\end{equation}
with\begin{equation}
A(\alpha) = \coth \alpha, 
\end{equation}
and
\begin{equation}
B(\alpha) =  \frac{\alpha^2 \coth \alpha -2\alpha + \sinh \alpha \cosh \alpha}{\sinh^2 \alpha - \alpha^2}.
\end{equation}
where $\boldsymbol{k}$ is the two-dimensional wavenumber with magnitude $k$ and unit vector $\hat{\boldsymbol{k}}$.
If the surface velocity further satisfies $\nabla\bcdot \boldsymbol{U} = 0$ {\it everywhere} on $z=0$, as has been assumed in other works (\emph{e.g.} \cite{Alexander_etal2007},~\cite{LubenskyGoldstein1996}, \cite{StoneMcConnell1995}), Eqn.~(\ref{fteqn}) then simplifies dramatically to
\begin{equation}
\hat{\boldsymbol{f}} = \mu k A(kH) \hat{\boldsymbol{U}} = \mu k \coth (kH)\hat{\boldsymbol{U}}.
\label{FTcond}
\end{equation}
Such an assumption confers the advantage of making $w=0$ and $p=0$ in the bulk fluid, which in turn makes some problems analytically tractable.
However, we emphasize that this simplification does not follow from assuming that
$\nabla\bcdot \boldsymbol{U} = 0$ in $\mathcal{D}$ alone, even in the axisymmetric case, so we will not make that assumption here. 

Equation~(\ref{fteqn}) can be rewritten as
\begin{equation}
    \hat{\boldsymbol{U}} =\frac{1}{\mu k} \left[ \frac{1}{A(kH)} \mathsfbi{I} - \frac{B(kH)}{A(kH)[A(kH) + B(kH)]} \hat{\boldsymbol{k}} \hat{{\boldsymbol{k}}} \right] \hat{\boldsymbol{f}},
    \label{Uhateqn}
\end{equation}
which can be interpreted as a statement that $\boldsymbol{U}$ is a convolution of $\boldsymbol{f}$ against some tensorial Green's function $\mathsfbi{G_H}$, expressible as an inverse 2D Fourier transform:
\begin{equation}
    \mathsfbi{G_H} = \frac{1}{(2\pi)^2} \int_0^{2\pi} \mathrm{d}\phi \int_0^\infty \mathrm{d}k~k \frac{1}{\mu k} \left[ \frac{1}{A(kH)} \mathsfbi{I} - \frac{B(kH)}{A(kH)[A(kH) + B(kH)]}\hat{\boldsymbol{k}} \hat{\boldsymbol{k}} \right] e^{\mathrm{i} kr\cos(\theta-\phi)}
\end{equation}
where $r$ and $\theta$ are the polar coordinates of real space and $k$ and $\phi$ are the polar coordinates of Fourier space. We define $\beta = \theta - \phi$, change variables, and apply periodicity in $\beta$ to obtain 
\begin{equation}
    \mathsfbi{G_H} = \frac{1}{(2\pi)^2} \int_0^{2\pi} \mathrm{d}\beta \int_0^\infty \mathrm{d}k~ \frac{1}{\mu}\left[\frac{1}{A(kH)}\mathsfbi{I} - \frac{B(kH)}{A(kH)[A(kH) + B(kH)]} \hat{\boldsymbol{Z}}\hat{\boldsymbol{Z}}\right]e^{\mathrm{i} kr\cos \beta}
\end{equation}
where $\hat{\boldsymbol{Z}} = \cos \beta \hat{\boldsymbol{x}} -\sin \beta \hat{\boldsymbol{x}}^\perp$ comes from expressing $\hat{\boldsymbol{k}}$ in the  $\{\hat{\boldsymbol{x}},\hat{\boldsymbol{x}}^\perp\}$ basis. Performing a Jacobi-Anger expansion (Formula 8.511.1 of  \cite{GradshteynRyzhik2007}) to evaluate the integrals over $\beta$, we arrive at
\begin{equation}
   \mathsfbi{G_H} = \frac{1}{2\pi \mu} \int_0^\infty \mathrm{d}k~\left\{\frac{J_0(kr)}{A(kH)}\mathsfbi{I}-\frac{B(kH)}{A(kH)[A(kH)+B(kH)]} \left[{J_1'(kr)}\hat{\boldsymbol{x}}\hat{\boldsymbol{x}} + \frac{J_1(kr)}{kr} \hat{\boldsymbol{x}}^\perp \hat{\boldsymbol{x}}^\perp\right]\right\},
   \label{Ghfull}
\end{equation}
where $J_0$ and $J_1$ are Bessel functions of the first kind. 
By the convolution theorem and Eqn.~(\ref{momentumeqnchi}),
\begin{equation}
    \boldsymbol{U}=\mathsfbi{G_H}*[\chi(\mathcal{D})\bnabla\bcdot\boldsymbol{\sigma}]
    =\mathsfbi{G_H}*[\chi(\mathcal{D})(-\nabla P + \bar\eta \bnabla^2 \boldsymbol{U})].
    \label{convolutioneqn}
\end{equation}
This equation holds for all points in the $z=0$ plane and, if evaluated inside of $\mathcal{D}$, generates an integral relation for the unknown $\boldsymbol{U}$. No standard method exists for its solution. The first expression involving
$\bnabla\bcdot\boldsymbol{\sigma}$ serves only to emphasize its generality -- other surface stress tensors could be considered. For example, if $\boldsymbol{\sigma}$ is Newtonian, Eqn.~(\ref{convolutioneqn}) could model a classical incompressible Langmuir film~(see \cite{Alexander_etal2006}). Equation~(\ref{convolutioneqn}) is likewise applicable to other materials such as active nematic films (e.g. \cite{GBGBS2015a}).

Thus, Eqn.~(\ref{convolutioneqn}), together with the incompressibility condition~(\ref{divUeqn}), the stress boundary condition (\ref{stressbcedge}), and the kinematic boundary condition, $\boldsymbol{V}=\boldsymbol{U}|_{\boldsymbol{x}\in\partial\mathcal{D}}$, completely specifies the initial value problem for $\partial\mathcal{D}$. Note Eqn.~(\ref{convolutioneqn}) is easily extended to a domain composed of multiple monolayers.

There are two limits where $\mathsfbi{G_H}$ takes a particularly simple closed form. 

{\bf (i)} In the limit of infinite $H$, $A(kH)$ and $B(kH)$ both approach unity with zero slope, and $\mathsfbi{G_H}$ approaches
\begin{equation}
    \mathsfbi{G_\infty} = \frac{1}{2\pi \mu r} \left(\mathsfbi{I} - \frac{1}{2} {\hat{\boldsymbol{x}}^\perp\hat{\boldsymbol{x}}^\perp}\right).
\end{equation}
Using the fact that $\mathsfbi{I} = \hat{\boldsymbol{x}}\hat{\boldsymbol{x}}+\hat{\boldsymbol{x}}^\perp\hat{\boldsymbol{x}}^\perp$, we can rewrite this expression as
\begin{equation}
    \mathsfbi{G_\infty} =  \frac{\mathsfbi{I} + \hat{\boldsymbol{x}}\hat{\boldsymbol{x}}}{4\pi \mu r}.
\end{equation}
That is, in the case of infinite subphase depth, the Green's function is proportional to the classical Stokeslet.

{\bf (ii)} In the limit of small depth, $H\to 0$, $A(kH) \to (kH)^{-1}$ and $B(kH) \to (kH/3)^{-1}$. In this case, we interpret the formal limit,  
\begin{equation}
    \mathsfbi{G_0}=\frac{H}{2\pi \mu} \int_0^\infty \mathrm{d}k~  \left\{kJ_0(kr)\mathsfbi{I} - \frac{3}{4}k \left[J_1'(kr) \hat{\boldsymbol{x}}\hat{\boldsymbol{x}} + \frac{J_1(kr)}{kr}\hat{\boldsymbol{x}}^\perp\hat{\boldsymbol{x}}^\perp\right]\right\}~,
\end{equation}
using generalized functions. Starting with the orthogonality relation (Formula 6.512.8 of~\cite{GradshteynRyzhik2007})
\begin{equation}
    \int_0^\infty \mathrm{d}k~kJ_0(kr)J_0(kr') = \frac{\delta(r-r')}{r}
\end{equation}
and letting $r'\to 0$, we find that
\begin{equation}
    \int_0^\infty \mathrm{d}k~kJ_0(kr) = \frac{\delta(r)}{r}.
\end{equation}
This can be combined with the convergent integral
\begin{equation}
    \int_0^\infty \mathrm{d}k~ k\frac{J_1(kr)}{kr}=\frac{1}{r^2}
\end{equation}
to show
\begin{equation}
    \int_0^\infty \mathrm{d}k~kJ_1'(kr) = \int_0^\infty \mathrm{d}k~\left[J_0(kr)-\frac{J_1(kr)}{kr}\right] = \frac{\delta(r)}{r} - \frac{1}{r^2}.
\end{equation}
Thus,
\begin{equation}
    \mathsfbi{G_0}=  \frac{H}{2\pi \mu} \left\{\frac{\delta(r)}{r} \mathsfbi{I} - \frac{3}{4} \left[\left(\frac{\delta(r)}{r} - \frac{1}{r^2}\right)\hat{\boldsymbol{x}}\hat{\boldsymbol{x}} + \frac{1}{r^2} \hat{\boldsymbol{x}}^\perp\hat{\boldsymbol{x}}^\perp\right]\right\}.
    \label{G0eqn}
\end{equation}
We can rewrite the quantity in square brackets by observing that its Fourier transform upon convolution with $\boldsymbol{f}$ is
\begin{equation}
     2\pi\left(\hat{\boldsymbol{k}}\hat{\boldsymbol{k}}\right) \cdot \hat{\boldsymbol{f}} = -2\pi\mathrm{i}\frac{\hat{\boldsymbol{k}}}{k} \left(\mathrm{i} {\boldsymbol{k}}\cdot \hat{\boldsymbol{f}}\right).
     \label{assoceqn}
\end{equation}
Since $-1/k^2$ is the Fourier transform of the fundamental solution to the Laplace equation, $-\mathrm{i}\hat{\boldsymbol{k}}/k$ is the Fourier transform of its gradient, namely $\hat{\boldsymbol{x}}/(2\pi r)$ in two dimensions. Applying the convolution theorem to the quantity on the RHS of Eqn.~(\ref{assoceqn}) thus establishes the relation
\begin{equation}
    \left[\left(\frac{\delta(r)}{r} - \frac{1}{r^2}\right)\hat{\boldsymbol{x}}\hat{\boldsymbol{x}} + \frac{1}{r^2} \hat{\boldsymbol{x}}^\perp\hat{\boldsymbol{x}}^\perp \right]*\boldsymbol{f}= \frac{\hat{\boldsymbol{x}}}{r} * \left(\bnabla \bcdot {\boldsymbol{f}}\right),
\end{equation}
so that
\begin{equation}
    \boldsymbol{U} = \mathsfbi{G_0} * \boldsymbol{f}= \frac{H}{2\pi \mu} \left[\frac{\delta(r)}{r}\mathsfbi{I} * \boldsymbol{f} - \frac{3}{4} \frac{\hat{\boldsymbol{x}}}{r}* (\bnabla\bcdot \boldsymbol{f})\right]=\frac{H}{\mu}\left[ \boldsymbol{f} - \frac{3}{4} \frac{\hat{\boldsymbol{x}}}{2\pi r} * (\bnabla\bcdot \boldsymbol{f})\right].
\end{equation}
Taking a divergence and using the fact that $\bnabla \bcdot (\hat{\boldsymbol{x}}/r) = 2\pi \delta_2(\boldsymbol{x})$, 
{where $\delta_2(\boldsymbol{x})$ is the two-dimensional Dirac delta distribution}, we arrive at the simple relation
\begin{equation}
    \bnabla\bcdot \boldsymbol{U} = \frac{H}{4\mu}(\bnabla \bcdot \boldsymbol{f}).
    \label{smallHdiv}
\end{equation}
Note that this result is also directly obtainable by letting $H\to 0$ in Eqn.~(\ref{fteqn}), which gives
\begin{equation}
    \hat{\boldsymbol{f}} = \frac{\mu}{H} \left(\mathsfbi{I} + 3\hat{\boldsymbol{k}}\hat{\boldsymbol{k}}\right) \hat{\boldsymbol{U}},
\end{equation}
multiplying by $\mathrm{i}\hat{\boldsymbol{k}}$, and taking an inverse Fourier transform.

Equation~(\ref{smallHdiv}), in addition to the assumptions that $\bnabla\bcdot \boldsymbol{U}= 0$ in $\mathcal{D}$ and $\boldsymbol{f} = 0$ in $\mathcal{D}^C$, is sufficient to guarantee that $\bnabla \bcdot \boldsymbol{U} = \bnabla \bcdot \boldsymbol{f}= 0$ on the entire surface so that ultimately
\begin{equation}
    \mathsfbi{G_0}=\frac{\delta_2(\boldsymbol{x})\mathsfbi{I}}{ \Gamma },
\end{equation}
where $\Gamma = \mu/H$. Equation~(\ref{convolutioneqn}) then becomes
\begin{equation}
   \chi(\mathcal{D})(-\nabla P + \bar\eta \bnabla^2\boldsymbol{U}) =\Gamma\boldsymbol{U}.
   \label{BrinkmanLimit}
\end{equation}
That is, at leading order, the forcing is not only linear but local in $\boldsymbol{U}$, and the equation of motion reduces to a Brinkman equation. Consequently, the pressure inside the monolayer is harmonic, and the fluid velocity outside of the monolayer is zero. Expanding Eqn.~(\ref{FTcond}) further in small $H$ yields 
\begin{equation}
    \hat{\boldsymbol{f}} = \mu k \left(\frac{1}{kH} + \frac{kH}{3} + \hdots\right)\hat{\boldsymbol{U}} = \Gamma \hat{\boldsymbol{U}} + \frac{\mu H}{3} k^2 \hat{\boldsymbol{U}} + \hdots
\end{equation}
Thus, in real space the $O(H)$ correction enters as a term proportional to $-\Delta \boldsymbol{U}$, which has the effect of slightly increasing the monolayer viscosity.

In summary, the small $H$ limit of the low friction case is the high friction case with substrate drag coefficient $\Gamma$. 



\section{The axisymmetric case}
\label{sec:physsystem}


\begin{figure}
    \centering
    \includegraphics[scale=0.165]{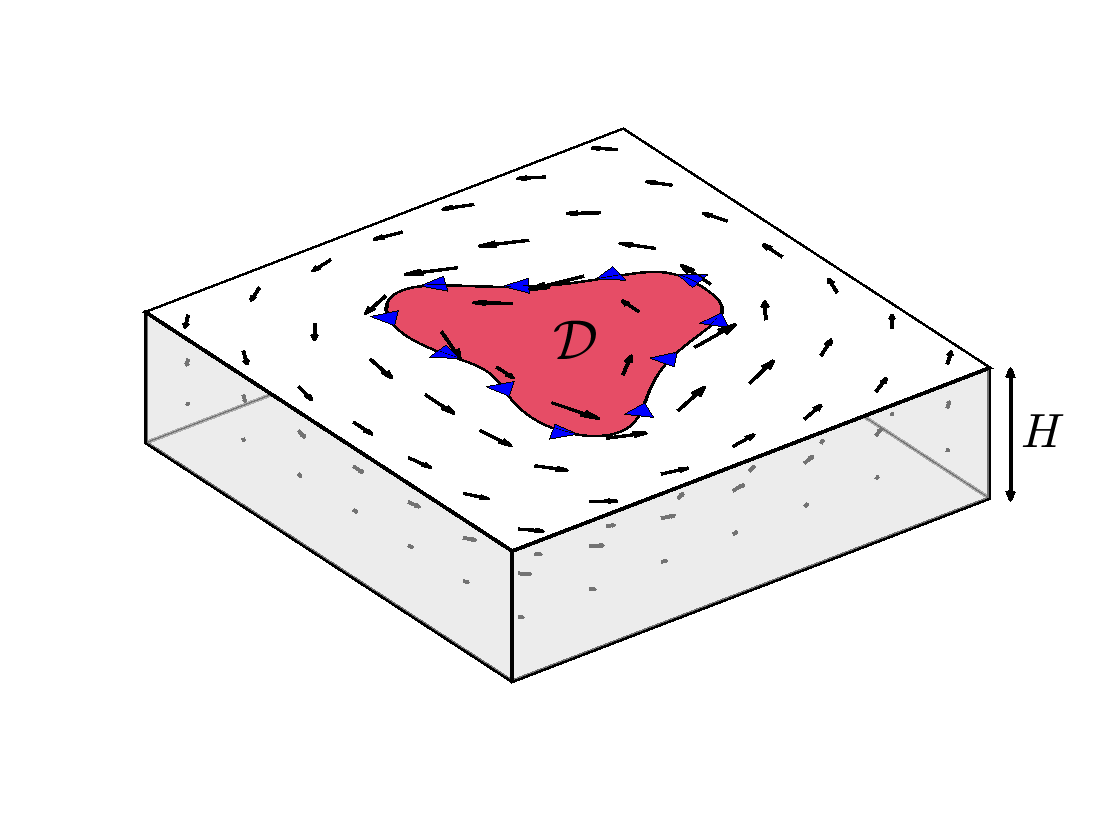}
    \includegraphics[scale=0.165]{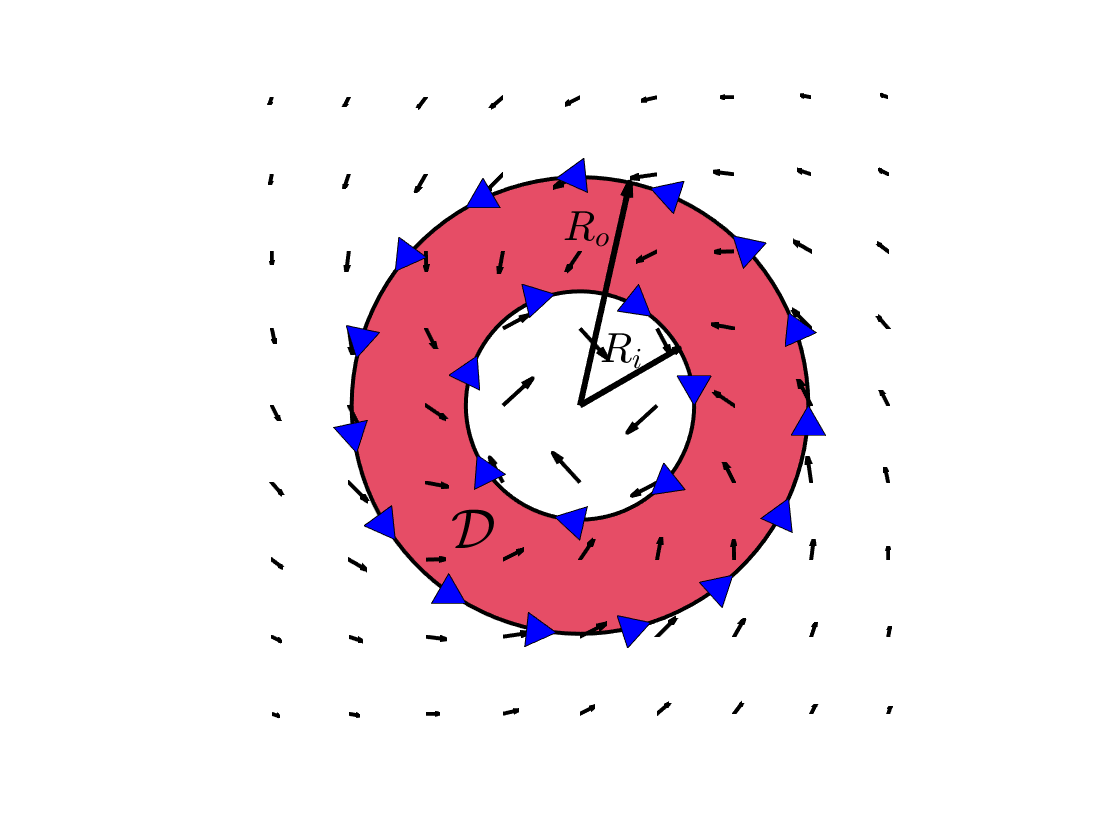}
    \caption{(Left) Schematic of a 2D monolayer domain $\mathcal{D}$ situated on the upper surface of a Stokesean sublayer of depth $H$, whose motion generates a shear stress on the monolayer. Outside of $\mathcal{D}$, the surface is stress-free. The inhomogeneous forcing arises from line tension as well as rotational stresses at the boundary of $\mathcal{D}$, represented by colored triangles. (Right) Top-down view of an annular domain with inner radius $R_i$ and outer radius $R_o$ and the corresponding axisymmetric surface flow field $\boldsymbol{U}$. Note that the rotational traction vectors follow the direction of the local tangent vector.}
    \label{fig:expcavity}
\end{figure}


We proceed to specialize the integral equation derived above to axisymmetric domains of either disks or annuli. It is advantageous to consider the radial and azimuthal equations separately. To this end, let $\boldsymbol{U} = U(r) \hat{\boldsymbol{x}} + V(r)\hat{\boldsymbol{x}}^\perp$ and $\boldsymbol{f} = f(r)\hat{\boldsymbol{x}} + g(r) \hat{\boldsymbol{x}}^\perp$. 
Substituting into Eqn.~(\ref{momentumeqn}), we obtain the momentum equations
\begin{equation}
   \chi(\mathcal{D}) \left(-\frac{\mathrm{d} P}{\mathrm{d} r} + \bar\eta \mathcal{L}[U] \right)= f
\end{equation}
\begin{equation}
   \chi(\mathcal{D}) \bar\eta{\mathcal{L}}[V] = g,
\end{equation}
where we have defined the differential operator
\begin{equation}
    \mathcal{L} = \frac{\partial^2}{\partial r^2} + \frac{1}{r} \frac{\p }{\partial r} - \frac{1}{r^2}.
    \label{Ldef}
\end{equation}

For axisymmetric quantities, 2D Fourier transforms reduce to Hankel transforms:
\begin{equation}
    \hat{\boldsymbol{U}} = -2\upi \mathrm{i}\hat{\boldsymbol{k}} \int_0^\infty \mathrm{d}r~ r U(r) J_1(kr) - 2\upi \mathrm{i} \hat{\boldsymbol{k}}^\perp\int_0^\infty \mathrm{d}r~r V(r) J_1(kr)
\end{equation}
\begin{equation}
    \hat{\boldsymbol{f}} = -2\upi \mathrm{i}\hat{\boldsymbol{k}} \int_0^\infty \mathrm{d}r~ r f(r) J_1(kr) - 2\upi \mathrm{i} \hat{\boldsymbol{k}}^\perp\int_0^\infty \mathrm{d}r~r g(r) J_1(kr).
\end{equation}
Substituting these expressions into Eqn.~(\ref{Uhateqn}) yields the relations
\begin{equation}
\int_0^\infty \mathrm{d}r~r U(r) J_1(kr) = \frac{1}{\mu k [A(kH) + B(kH)]} \int_0^\infty \mathrm{d}r~rf(r) J_1(kr)
\label{ftradial}
\end{equation}
\begin{equation}
\int_0^\infty \mathrm{d}r~rV(r) J_1(kr) = \frac{1}{\mu k A(kH) }\int_0^\infty \mathrm{d}r~rg(r) J_1(kr).
\label{fttangential}
\end{equation}
These equations show that the azimuthal and radial dynamics are decoupled in the bulk, but as we will see, this is not always the case at the boundary $\partial \mathcal{D}$, where the two interact through the odd viscosity. By taking another Hankel transform to solve for the velocity components, we obtain the axisymmetric analogues of Eqn.~(\ref{convolutioneqn}):
\begin{equation}
    U(r) = \frac{1}{\mu}\int_0^\infty \mathrm{d}k~ \frac{J_1(kr)}{A(kH)+B(kH)} \int  \mathrm{d}r'~r'f(r') J_1(kr')= \frac{1}{\mu} \int_{R_i}^{R_o} \mathrm{d}r'~r' M(r,r') f(r')
    \label{Usie}
\end{equation}
\begin{equation}
    V(r) = \frac{1}{\mu}\int_0^\infty \mathrm{d}k~ \frac{J_1(kr)}{A(kH)} \int \mathrm{d}r'~r'g(r') J_1(kr')=\frac{1}{\mu} \int_{R_i}^{R_o} \mathrm{d}r'~r' L(r,r') g(r')
    \label{Vsie}
\end{equation}
where $R_i<R_o$, $R_i$ can be zero, and $R_o$ can be positive infinity. We have also defined the kernels
\begin{equation}
    M(r,r') = \int_0^\infty \mathrm{d}k~\frac{J_1(kr)J_1(kr')}{A(kH) + B(kH)}
\end{equation}
\begin{equation}
    L(r,r') = \int_0^\infty \mathrm{d}k~\frac{J_1(kr)J_1(kr')}{A(kH)}.
\end{equation}

In the limit $H\to\infty$, we obtain closed form expressions for $M$ and $L$, which we denote with a bar, from \cite{GradshteynRyzhik2007} Formulas 6.512.1, 8.126.3, and 8.126.4:
\begin{align}
\bar{L}(r,r')= 2\bar{M}(r,r') &= \int_0^\infty \mathrm{d}k~ J_1(kr)J_1(kr') \\
&=  \left\{\begin{array}{cc}
\frac{2}{\pi r}(-E[r^2/r'^2] + K[r^2/r'^2]) & \text{ if }r< r'\\
\frac{2}{\pi r'}(-E[r'^2/r^2] + K[r'^2/r^2]) & \text{ if }r'<r
\end{array}\right.\\
&=-\frac{1}{\pi rr'(r+r')}\left[(r+r')^2E[\xi] -(r^2+r'^2)K[\xi]\right]
\end{align}
where $\xi = 4rr'/(r+r')^2$, and $K$ and $E$ are complete elliptic integrals of the first and second kind, respectively. The reader is cautioned that our notational convention for elliptic integrals differs from that of \cite{GradshteynRyzhik2007} by a square root in the argument.  Since $K$ has a logarithmic singularity when its argument approaches unity, these expressions illustrate that $L(r,r')$ and $M(r,r')$ are both logarithmically singular when $r\sim r'$, and it becomes useful to isolate the most singular behavior. We write
\begin{equation}
    \bar{L}(r,r') = -\frac{\log|r-r'|}{\pi r'} + \tilde{L}(r,r')
\end{equation}
where $\tilde{L}$ has a removable singularity at $r=r'$, with an analogous expansion for $\bar{M}$.

In the opposite limit of $H\to 0$, $A(kH) \sim (kH)^{-1}$, and $L$ is proportional to a Dirac delta:
\begin{equation}
    \int_0^\infty \mathrm{d}k~ kH J_1(kr)J_1(kr') = H \frac{\delta(r-r')}{r'},
\end{equation}
using orthogonality properties of $J_1$ (Formula 6.512.8 from \cite{GradshteynRyzhik2007}). The convolution integral thus reduces to a Brinkman ordinary differential equation  whose solution is discussed in \textsection\ref{smallHlimit} and \textsection\ref{smallHsol_ann}. For the radial direction, we similarly find $M\to 4H\delta(r-r')/r'$.

As noted by \cite{YanSloan1988}, integral equations with logarithmically singular kernels generally have unique solutions 
that diverge like an inverse square root of the distance to the boundary.
For our domains, we can therefore expect the surface shear stresses of the bulk fluid $f(r)$ and $g(r)$ to take the form
\begin{equation}
    f(r) = \frac{\tilde{f}(r)}{\sqrt{r-R_i}\sqrt{R_o-r}}\chi(R_i<r<R_o)
\end{equation}
and
\begin{equation}
    g(r) = \frac{\tilde{g}(r)}{\sqrt{r-R_i}\sqrt{R_o-r}}\chi(R_i<r<R_o)
\end{equation}
for smooth functions $\tilde{f}$ and $\tilde{g}$ that do not vanish at $r=R_i$ and $r=R_o$.
As such, $V''(r)$, as well as $P'(r)$ in the case of an annulus, similarly diverge like an inverse square root of the distance to the boundaries when the limit is taken from the domain interior. (The exterior velocity fields will also exhibit singularities in their derivatives as the boundary is approached; see \textsection 5.2 for one analytical example.) Nonetheless, in spite of these divergences, both $V(r)$ and $V'(r)$ remain bounded. This phenomenon is consistent with other axisymmetric systems with vanishingly thin domains immersed in a continuous three-dimensional medium, such as the rotating solid disk submerged in a Stokes fluid analyzed by~\cite{Jeffery1915} or the penny-shaped crack in a three-dimensional elastic medium analyzed  by~\cite{Sneddon1946}. For the rotating disk, \cite{Sherwood2013} has found that relaxing the no-slip condition by allowing for a finite slip length can regularize this singularity, but such a condition is not present in our model. 
A consequence of the divergence at the boundary is that even the problem of the linearly perturbed disc requires a great deal of subtlety, as linearization requires further differentiation of the boundary terms.
Linear stability analysis of a related system was previously treated by~\cite{StoneMcConnell1995}, although they included a simplifying global incompressibility assumption as well as a nonzero surface viscosity everywhere, which circumvents the issue at hand. A linear stability analysis for the $H\to 0$ case was previously completed  in~\cite{Soni_etal2019} since the integral kernel is delta singular rather than logarithmically singular and velocity gradients do not diverge in this case.

In the following sections, we describe several ways to derive and numerically solve the singular integral equations Eqns.~(\ref{Usie}) and~(\ref{Vsie}) for two important axisymmetric geometries: the disc and the annulus.


\section{A disc-shaped domain}
\label{sec:disc}


We begin with the simplest nontrivial axisymmetric geometry and take the domain $\mathcal{D}$ to be the disc of radius $R$ centered at the origin. The incompressibility of the monolayer disc centered at the origin automatically implies that there is no radial component to the axisymmetric flow; consequently, the entire surface flow field is incompressible and Eqn.~(\ref{FTcond}) holds. The pressure gradient also vanishes. Hence, we can formulate the problem of finding the flow field on the surface as a scalar mixed boundary value problem using Eqn.~(\ref{momentumeqnchi}). Let $v(r,z)$ be the azimuthal component of $\boldsymbol{u}(r,z)$ and let $V(r) = v(r,z=0)$. 
The azimuthal momentum equation can be written as
\begin{equation}
\left. \begin{array}{ll}  
\displaystyle\left. \mu \frac{\partial v}{\partial z}\right|_{z=0}=\bar\eta \left(\frac{\mathrm{d}^2 V}{\mathrm{d} r^2} + \frac{1}{r} \frac{\mathrm{d} V}{\mathrm{d} r} - \frac{V}{r^2} \right),
  \quad \mbox{for\ }\quad r<R,\\[8pt]
\displaystyle\left. \mu \frac{\partial v}{\partial z}\right|_{z=0}= 0,
  \quad \mbox{for\ }\quad r>R,
 \end{array}\right\}
  \label{dropleteqns}
\end{equation}
or, more compactly,
\begin{equation}
    g(r) = \bar\eta \mathcal{L}[V(r)]\chi(r<R),
\end{equation}
where the 2D vector Laplacian operator  $\mathcal{L}$ was defined in Eqn.~(\ref{Ldef}). If the 2D surface pressure outside of the monolayer is taken to be zero, the stress boundary condition Eqn.~(\ref{stressbcedge}) in this geometry reduces to the Robin boundary condition
\begin{equation}
V'(R^-) - \frac{\eta-\eta_R}{\bar\eta R} V(R) = \frac{2\eta_R\Omega}{\bar\eta}
\label{robinbc}
\end{equation}
in the azimuthal direction; it is unnecessary to analyze the radial direction as it ultimately simply sets the pressure difference. Note that in this situation, the absence of a radial velocity means the odd stress only produces a transverse stress which is balanced by the pressure. Also note this boundary condition is what introduces an inhomogeneous forcing into the problem. We demonstrate two techniques to solve for the droplet flow field: a direct inversion of the singular integral equation in convolution form Eqn.~(\ref{Vsie}) and an equivalent formulation as a dual integral equation for the Hankel transformed velocity field that admits a semi-analytic solution.

\subsection{Solution via Green's function.}

We begin by noting that the solution to Eqn.~(\ref{dropleteqns}) for $r<R$ can be written as
\begin{equation}
V(r) = V_p(r) + V_h(r),
\end{equation}
where the particular solution is of the form
\begin{equation}
V_p(r) = \int \mathrm{d}r' ~G(r,r') g(r'),
\end{equation}
with the Green's function $G$ of the operator $\bar\eta\mathcal{L}$ satisfying
\begin{equation}
\bar\eta  \left( \frac{\partial^2}{\partial r^2} + \frac{1}{r} \frac{\partial }{ \partial r} - \frac{1}{r^2}\right) {G(r,r')} = \delta(r-r'),
\label{GFeqn}
\end{equation}
and the homogeneous solution is of the form
\begin{equation}
V_h(r) = A_hr + \frac{B_h}{r},
\end{equation}
with constants $A_h$ and $B_h$ that are chosen to satisfy the boundary condition Eqn.~(\ref{robinbc}) as well as the condition $V(0) = 0$ that is a consequence of axisymmetry. 

The Green's function $G(r,r')$ is easily calculated via Hankel transforms. We find
\begin{equation}
G(r,r') = -\frac{1}{\bar\eta} \int_0^\infty \mathrm{d}k ~\frac{J_1(kr)J_1(kr')}{k} = -\frac{1}{\bar\eta} \left[\frac{r}{2} \chi({r<r'}) + \frac{r'^2}{2r} \chi({r'<r})\right]
\end{equation}
so that
\begin{equation}
V_p(r) = \int_0^R \mathrm{d}r'~ G(r,r') g(r') = - \frac{1}{2\bar\eta r} \int_0^r \mathrm{d}r'~(r')^2 g(r') - \frac{r}{2\bar\eta} \int_r^R \mathrm{d}r' ~g(r').
\label{Vpdef}
\end{equation}
 It remains to solve for $A_h$ and $B_h$. The condition $V(0)=0$ implies that $B_h=0$. The azimuthal velocity $V= V_p + A_hr$ is continuous and piecewise smooth, so upon substitution into Eqn.~(\ref{robinbc}), we obtain
\begin{equation}
A_h = \Omega + \frac{\eta-\eta_R}{2\eta_R R} V_p(R) - \frac{\bar\eta}{2\eta_R} V_p'(R^-),
\end{equation}
where $R^-$ denotes the limit as $r$ approaches $R$ from the left. Using Eqn.~(\ref{Vpdef}), this can be written as
\begin{equation}
    A_h = \Omega - \frac{\eta}{2\eta_R \bar\eta R^2} \int_0^R \mathrm{d}r'~(r')^2 g(r').
    \label{Ahdef}
\end{equation}
Finally, substituting everything into Eqn.~(\ref{Vsie}) yields the following integral equation for $g$ in the interval $0<r<R$:
\begin{equation}
    \int_0^R \mathrm{d}r'~G(r,r') g(r') + r \left[\Omega - \frac{\eta}{2\eta_R \bar\eta R^2} \int_0^R \mathrm{d}r'~(r')^2 g(r')\right] = \frac{1}{\mu} \int_0^R \mathrm{d}r'~ r' L(r,r') g(r').
    \label{gie_disc}
\end{equation}
 The discretization of the singular kernel $L(r,r')$ is handled as follows. First, we expand in $H$ so that
 \begin{equation}
     L(r,r') = \bar{L}(r,r') + L^*(r,r')
 \end{equation}
 where 
 \begin{equation}
     L^*(r,r') = \int_0^\infty \mathrm{d}k~ J_1(kr)J_1(kr') (\tanh kH - 1).
 \end{equation}
Because the integrand of $L^*$ decays to zero exponentially, this integral is well-approximated by taking a finite cutoff, say by replacing the upper limit of $\infty$ by $10/H$, and numerically integrating. Next, we focus on the diagonal terms by writing the integral as
\begin{equation}
    \frac{1}{\mu} \int_0^R \mathrm{d}r'~r' \bar{L}(r,r') g(r') = \frac{1}{\mu}\int_0^R \mathrm{d}r' ~r'\bar{L}(r,r') [g(r')-g(r)] + \frac{1}{\mu}g(r)\int_0^R \mathrm{d}r'~r' \bar{L}(r,r').
\end{equation}
The first term on the RHS vanishes as $r$ approaches $r'$ if $g$ is smooth enough. Formula 6.561.13 from~\cite{GradshteynRyzhik2007} allows us to evaluate the second integral on the RHS, which we denote $L_d$, in terms of hypergeometric functions ${}_pF_q$:
\begin{align}
    L_d(r) &= \int_0^R \mathrm{d}r'~r'\bar{L}(r,r') \nonumber \\
    &= \int_0^\infty \mathrm{d}k~ J_1(kr) \int_0^R\mathrm{d}r'~r'J_1(kr') \nonumber \\
    &=\frac{r}{32}\left(-8 - \frac{3r^2}{R^2}~{}_4F_3\left[\{1,1,\frac{3}{2},\frac{5}{2}\},\{2,2,3\},\frac{r^2}{R^2}\right]+ 16 \log \frac{4R}{r}\right).
\end{align}
Once $g$ is known, Eqns.~(\ref{Vpdef}) and~(\ref{Ahdef}) can be used to calculate $V$ inside the monolayer. When $r>R$, the integral equation is nonsingular away from $r=R$ and the same $g$ can be substituted into Eqn.~(\ref{Vsie}) to calculate $V$ directly. 

\begin{figure}
\centering
\includegraphics[scale=0.16]{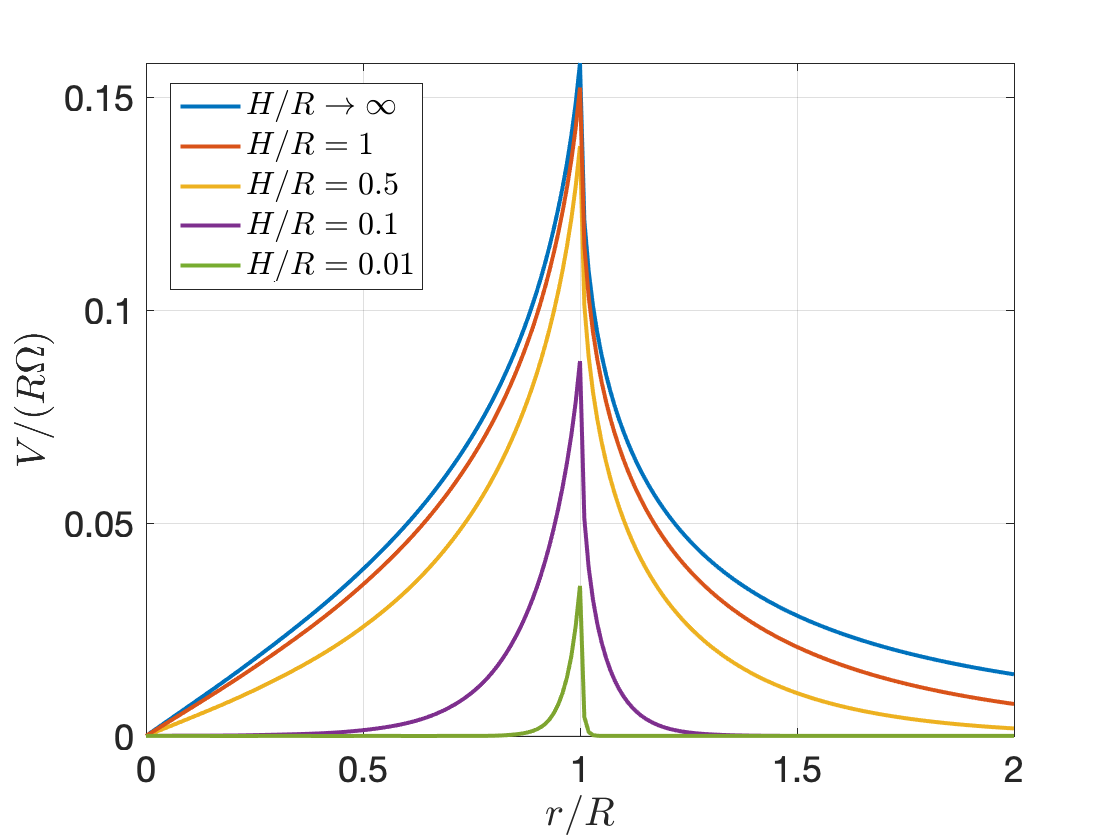}
\includegraphics[scale=0.16]{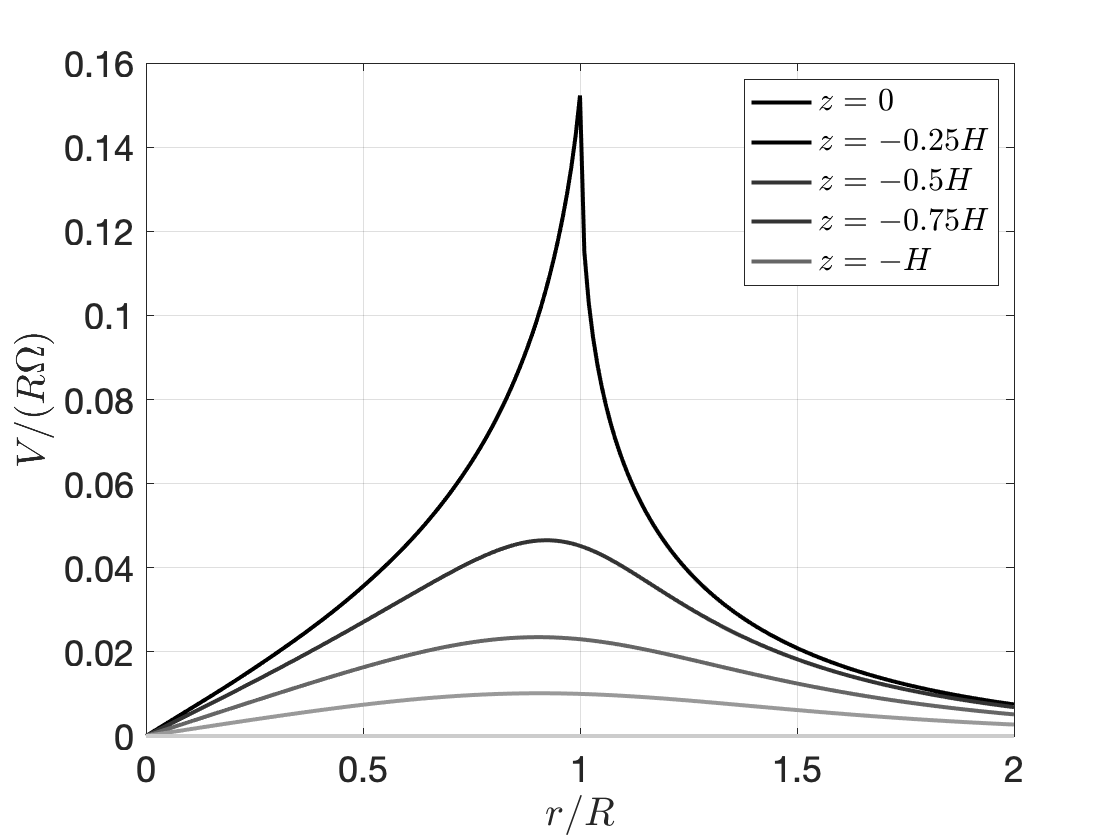}
\caption{(Left) The surface azimuthal velocity $V$ across an active disc-shaped monolayer of radius $R$ as subphase depth $H$ is varied. As $H\to 0$, a boundary layer of width $\bar{\delta} = \sqrt{\bar\eta H/\mu}$ becomes visible and $V$ is well-approximated by Eqn.~(\ref{smallHsol}). 
Parameters: $\eta_R/\eta =1.875 $, $\mu R/\eta = 30$. Here, $R$ is large compared to the Saffman-Delbr\"uck length $\ell_{SD}=\eta/\mu$, and $V$ scales like $\ell_{SD}\Omega$ for fixed $H$. 
Since the radial velocity $U=0$ for the disc, the flow fields are independent of line tension and odd viscosity. (Right) For the same parameters, with $H/R=1$, the azimuthal velocity at different sublayer depths.}
\label{tvimg}
\end{figure}

Figure~\ref{tvimg}(left) shows $V(r)$ calculated for various subphase thicknesses $H$. Within $\mathcal{D}$, $V(r)$ is upwardly convex and at larger values of $H$ (well-described by the infinite $H$ case) shows nearly solid-body rotation near the droplet center with a delocalized, faster edge current. This agrees qualitatively with the observations of \cite{Soni_etal2019}. We find that $V$ is continuous and smooth everywhere except at the interface $r=R$. As the limit is taken from either side of the interface, $V'$ is found to be finite-valued but $V''$ diverges like $|r-R|^{-1/2}$, as previously discussed.
The maximal $V$ is always found at $r=R$; outside of the disc, $V(r)$ decays exponentially with increasing $r$ if $H$ is finite. (For infinite $H$, the rate of decay is an inverse quadratic.) 
Decreasing $H$ has the effect of generally reducing the motion both in the monolayer bulk and in the exterior. In the limit of $H\ll R$, the motion becomes largely localized to a boundary layer at $r=R$ whose thickness scales with the penetration depth $\bar{\delta} = \sqrt{\bar\eta H/\mu}$, and $V$ is well-approximated by  Eqn.~(\ref{smallHsol}). This agrees with the analytical prediction of an edge current in the high friction case, as discussed in \textsection4.4 and in \cite{Soni_etal2019}.

The above results can be framed in terms of the Saffman-Delbr\"uck length, $\ell_{SD} = \eta/\mu$. Nondimensionalizing the monolayer momentum equation in the simplest case where $\eta = \eta_R$ and $H\to\infty$, we obtain
\begin{equation}
    \bar{\beta} \mathcal{L}[V] =\left.\frac{\partial v}{\partial z}\right|_{z=0}.
\end{equation}
where all lengths have been scaled by $R$ and $\bar{\beta} = 2\ell_{SD}/R$ is a dimensionless parameter formed from a ratio of the only remaining length scales. If $\bar{\beta} \gg 1$, then momentum is dissipated primarily through the monolayer, and $\mathcal{L}[V]$ is necessarily small. Taken with the boundary condition Eqn.~(\ref{robinbc}), this implies $V \approx \Omega r$ inside of the monolayer. Thus, we recover the rotating rigid disk in an infinite Stokes fluid analyzed by \cite{Jeffery1915}. In this case, the bulk shear stress at the surface is
\begin{equation}
    \left.\mu \frac{\partial v}{\partial z}\right|_{z=0} = \frac{4\mu\Omega}{\pi}\frac{r}{\sqrt{R^2-r^2}}\chi(r<R)
\end{equation}
and the surface velocity field is given by
\begin{equation}
    V(r) = \left\{\begin{array}{cc}
    \Omega r&, 0<r<R\\
    \frac{2\Omega}{\pi r}\left[r^2\sin^{-1}\left(\frac{R}{r}\right) - R\sqrt{r^2 -R^2}\right]&,r>R
    \end{array}\right. 
\end{equation}
so that the aforementioned regularity properties are all explicitly observable.

In the opposite limit of $\bar{\beta} \ll 1$, momentum is dissipated primarily through the bulk subphase, and the dimensionless velocity profile is nonlinear. Now, the dimensionless $(\partial v/\partial z)|_{z=0}$ scales like $\bar\beta$ so that the dimensional velocity scales like $\ell_{SD}\Omega$ in this regime. This is the $\bar\beta$ regime depicted in Fig.~\ref{tvimg}.


Once the surface velocity has been found, the subphase velocity and pressure can be calculated by specializing the formulas that appear in~\cite{MasoudShelley2014}. Define the Hankel transformed velocity component
\begin{equation}
    \mathcal{H}[U(r')](k) = \int_0^\infty \mathrm{d}r' ~r'J_1(kr')U(r')
\end{equation}
and similarly for $V(r')$. For axisymmetric geometries,
\begin{equation}
    v(r,z) = \int_0^\infty \mathrm{d}k~ k J_1(kr)\mathcal{H}[V(r')] \frac{\sinh k(z+H)}{\sinh kH}
    \label{vbulk}
\end{equation}
\begin{align}
    &u(r,z) = \int_0^\infty \mathrm{d}k~ k J_1(kr)\mathcal{H}[U(r')]\times \nonumber\\ &\left\{\frac{\sinh k(z+H)}{\sinh kH} + \frac{kz\sinh kH \coth k(z+H) + kH [(kH \coth kH-1)\sinh kz - kz\cosh kz]}{\sinh^2 kH - k^2H^2}\right\} 
    \label{ubulk}
\end{align}
\begin{equation}
    w(r,z) = \int_0^\infty \mathrm{d}k ~kJ_0(kr)\mathcal{H}[U(r')] \frac{k^2H (z+H) \sinh kz - kz \sinh kH \sinh k(z+H)}{\sinh^2 kH -k^2 H^2}
    \label{wbulk}
\end{equation}
\begin{equation}
    p(r,z) = 2\mu \int_0^\infty \mathrm{d}k~ k^2 J_0(kr) \mathcal{H}[U(r')] \frac{\sinh kH \sinh k(z+H) - kH \sinh kz}{\sinh^2 kH -k^2 H^2}
    \label{pbulk}.
\end{equation}
For the disc, $u$, $w$, and $p$ are clearly seen to vanish because $U=0$. 
The azimuthal velocity in the subphase $v$ is plotted for various depths $z$ in Fig.~\ref{tvimg}(right). It satisfies the no-slip boundary condition at $z=-H$ and the continuity condition $v=V$ at $z=0$. For $-H<z<0$,
the exponential decay of the integrand as $k\to\infty$ in Eqn.~(\ref{vbulk}) ensures that the subsurface velocity field is smooth.

\subsection{A formulation as dual integral equations}

As a check, and as an extension of more classical approaches, we now reframe the problem of finding $V$ as solving a pair of integral equations, one each on the adjoining intervals $(0,R)$ and $(R,\infty)$. Define $a(k)$ to be the Hankel transform of $g(r)$:
\begin{equation}
g(r) = \int_0^\infty \mathrm{d}k~ k a(k)J_1(kr),
\label{adef}
\end{equation}
which, by virtue of Eqn.~(\ref{fttangential}), implies
\begin{equation}
V(r) =\frac{1}{\mu}  \int_0^\infty \mathrm{d}k~ \tanh kHa(k) J_1(kr).
\end{equation}
Substitute into Eqn.~(\ref{dropleteqns}) to obtain
\begin{equation}
\int_0^\infty \mathrm{d}k~ k  \left(\frac{\bar\eta}{\mu}k \tanh kH + 1\right) a(k)J_1(kr) = 0,\quad 0<r<R\label{de1}
\end{equation}
\begin{equation}
\int_0^\infty \mathrm{d}k~ k a(k)J_1(kr)=0,\quad r>R\label{de2}.
\end{equation}

Equations~(\ref{de1}) and~(\ref{de2}) constitute a set of dual integral equations with Bessel-type kernel in the unknown $a(k)$ that is homogeneous for $r>R$. Several powerful methods, such as those developed by \cite{Busbridge1938}, \cite{Cooke1956}, or \cite{Sneddon1975}, have been introduced over the years in order to solve problems of this form. Perhaps the most computationally straightforward of these is the method of ~\cite{Tranter1954}, who found an explicit countable basis for $a(k)$ and cast the problem as an infinite linear system, thus reducing the problem of finding the azimuthal velocity field to the problem of finding a small number of coefficients.
Tranter's method has previously been used in works such as \cite{Stone1995}, \cite{HenleLevine2009}, and \cite{MartinSmith2011} to solve Cartesian or axisymmetric mixed boundary problems in fluid mechanics where the flow or a stress is prescribed on an inner region, but the procedure here requires a modification since no such information is provided in our system--in the problem at hand, the inhomogeneous forcing  comes from the boundary condition Eqn.~(\ref{robinbc}). In the following section, we show how to modify Tranter's procedure in such a way that the boundary data naturally enters into the problem.

\subsubsection{Solution via Tranter's method}

We begin by deriving the momentum equation in Fourier space.
In the case of an axisymmetric monolayer, 2D Fourier transforms reduce to Hankel transforms:
\begin{equation}
\hat{\boldsymbol{f}} = \int_\mathcal{D} \mathrm{d}A~  \bar\eta \Delta \boldsymbol{U}e^{-\mathrm{i} \boldsymbol{k} \bcdot\boldsymbol{x}} = -2\upi \mathrm{i} \hat{\boldsymbol{k}}^\perp\int_0^R \mathrm{d}r~ r  \bar\eta \mathcal{L}[V](r)J_1(kr)
\end{equation}
and
\begin{equation}
\hat{\boldsymbol{U}} = \int_\mathcal{D} \mathrm{d}A ~ \boldsymbol{U}e^{-\mathrm{i} \boldsymbol{k} \bcdot\boldsymbol{x}} = -2\upi \mathrm{i}\hat{\boldsymbol{k}}^\perp \int_0^\infty \mathrm{d}r~ r V(r)J_1(kr).
\end{equation}
Here, $\mathcal{L}$ is the differential operator defined in Eqn.~(\ref{Ldef}).
It is advantageous to integrate by parts, 
\begin{equation}
\hat{\boldsymbol{f}} = -2\pi \mathrm{i} \bar\eta \hat{\boldsymbol{k}}^\perp \left[V'(R^-) R J_1(kR) + V(R) [J_1(kR)-  kR J_0(kR)] -k^2\int_0^R \mathrm{d}r~ r   V(r)J_1(kr)\right],
\end{equation}
and substitute in the stress boundary condition Eqn.~(\ref{robinbc}), to arrive at the integral version of the inhomogeneous momentum equation in Fourier space,
\begin{equation}
-2\eta_R \Omega R J_1(kR) = V(R) [2\eta J_1(kR) - \bar\eta kR J_0(kR)] - \bar\eta k^2 \int_0^R \mathrm{d}r~ r V(r) J_1(kr) -  a(k).
\label{integralequation}
\end{equation}
Here we have defined $a(k)$ to be the Hankel transform of $g(r)$ as in Eqn.~(\ref{adef}).

\cite{Tranter1954} observed that without loss of generality, we may take our $a(k)$ to be of the form
\begin{equation}
a(k) = k^{-\beta} \sum_{n=0}^\infty a_n J_{2n+1+\beta}(kR)
\end{equation}
where $\beta> 0 $ is arbitrary and the coefficients $a_n$ are unknown; such a form for $a(k)$ automatically satisfies the condition $\boldsymbol{f} = 0$ for $r>R$. This expression is substituted into the integral equation and projected back onto the basis to yield an infinite system of linear equations for the $\{a_n\}$, which are then truncated and solved to obtain $V$. By Eqn.~(\ref{fttangential}) and the Hankel inversion theorem,
\begin{equation}
V(r) =\frac{1}{\mu} \int_0^\infty \mathrm{d}k~ a(k) \tanh kH J_1(kr) = \frac{1}{\mu}\sum_{n=0}^\infty a_n \int_0^\infty \mathrm{d}k~ k^{-\beta} \tanh kH J_{2n+1+\beta}(kR) J_1(kr).
\label{hankelinversion}
\end{equation}
Upon substitution into Eqn.~(\ref{integralequation}),
\begin{align}
&-2\eta_R \Omega R \mu J_1(kR) \\
&=  [2\eta J_1(kR) - \bar\eta kR J_0(kR)]\sum_{n=0}^\infty a_n \int_0^\infty \mathrm{d}k'~ k'^{-\beta} \tanh k'H J_{2n+1+\beta}(k'R) J_1(k'R) \nonumber\\
&- \bar\eta k^2 \sum_{n=0}^\infty a_n\int_0^R \mathrm{d}r~ r J_1(kr) \int_0^\infty \mathrm{d}k'~ k'^{-\beta} \tanh kH J_{2n+1+\beta}(k'R) J_1(k'r)\nonumber \\
&- \mu k^{-\beta} \sum_{n=0}^\infty  a_n J_{2n+1+\beta}(kR).\nonumber
\end{align}
We now multiply the equation by $k^{-1-\beta} J_{2m+1+\beta}(kR)$, where $m$ is a nonnegative integer, and integrate  $k$ from $0$ to $\infty$. This yields the infinite system of equations
\begin{equation}
-2\eta_R \Omega R \mu g_n = \sum_{m=0}^\infty a_m \left[(2\eta g_n - \bar\eta\xi_n) \Lambda_m - \bar\eta M_{mn} - \mu \Delta_{mn}\right]
\label{trantersystem}
\end{equation}
where, using formulas from \cite{GradshteynRyzhik2007},
\begin{equation}
g_n = \int_0^\infty \mathrm{d}k~ J_1(kR) J_{2n+1+\beta}(kR) k^{-1-\beta} = 
\frac{R^\beta}{2^{1+\beta} (1+\beta)!}\delta_{n0},
\end{equation}
\begin{equation}
\xi_n = \int_0^\infty \mathrm{d}k~k J_0(kR) J_{2n+1+\beta}(kR) k^{-\beta} = 0,
\end{equation}
\begin{equation}
\Lambda_n = \int_0^\infty \mathrm{d}k'~(k')^{-\beta} \tanh k'H J_{2n+1+\beta}(k'R) J_1(k'R) ,
\end{equation}
\begin{equation}
M_{mn} = \int_0^\infty \mathrm{d}k'~ (k')^{-2\beta}\tanh k' HJ_{2n+1+\beta}(k'R) J_{2m+1+\beta}(k'R),
\end{equation}
\begin{align}
\Delta_{mn} &= \int_0^\infty \mathrm{d}k~ k^{-1-2\beta} J_{2n+1+\beta}(kR)J_{2m+1+\beta}(kR) \nonumber \\
&= \frac{ \beta R^{2\beta}  (2\beta-1)! (m+n)!}{4^{\beta}(\beta+m-n)!(\beta-m+n)!(1+2\beta +m + n)!}.
\label{deltaeqn}
\end{align}
Here, noninteger factorials assume their usual definition via the gamma function (Formula 8.310.1 of \cite{GradshteynRyzhik2007}). In cases where a negative integer factorial is being taken in the denominator, the factorial is interpreted as infinity so that the integral vanishes.

At this stage, we choose $\beta=1/2$ so that the matrix $M_{mn}$ is nearly diagonal for large $k$. If $H\to\infty$, then $\beta=1/2$ ensures 
$M_{mn}$ is exactly diagonal. This choice of $\beta$ aids the convergence of the numerical routine by capturing the anticipated inverse square root divergence of $g(r)$ at the boundary and is hence optimal, although we emphasize that the routine converges to the same solution regardless of $\beta$.

\begin{figure}
    \centering
    \includegraphics[scale=0.2]{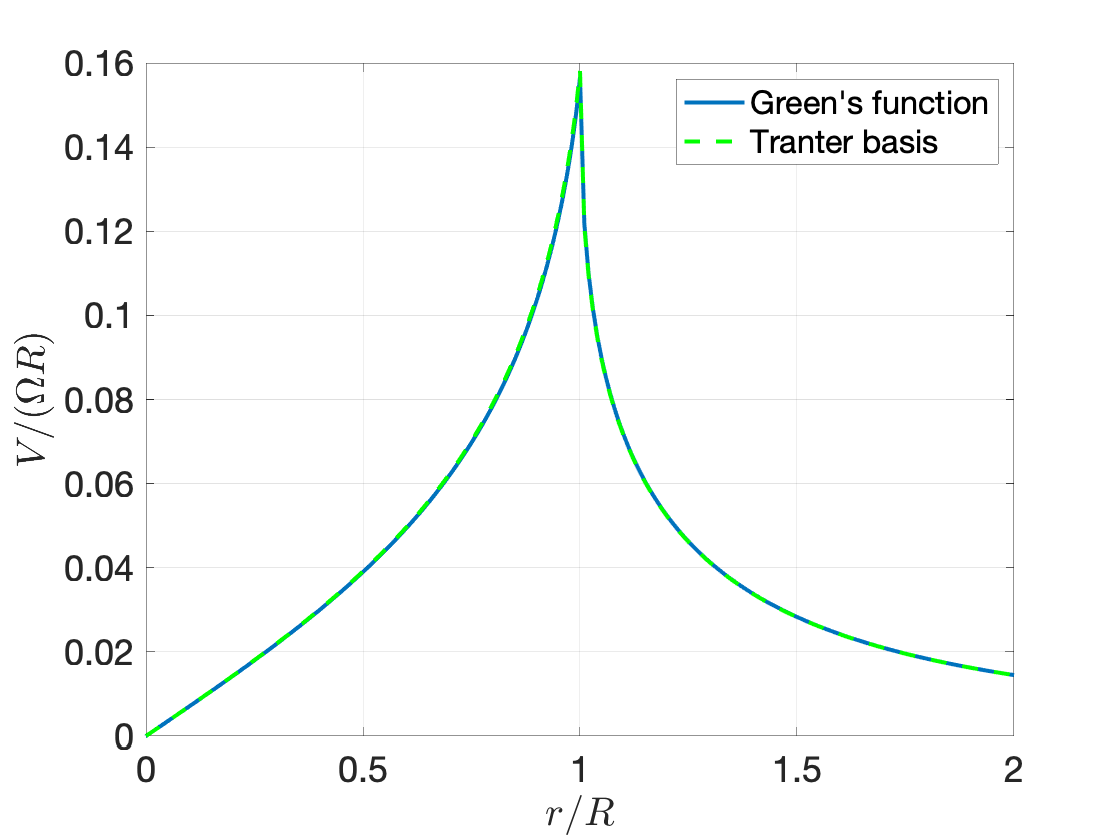}
    \caption{Comparison of the solution of the integral equation Eqn.~(\ref{Vsie}) computed using Tranter's method (20 terms, dashed line) with the solution using the Green's function formulation (800 Chebyshev nodes on the interval $0<r<R$, solid line) for a disc-shaped monolayer on a subphase of infinite depth. Parameters: $\eta_R/\eta = 1.875$, $\mu R/\eta = 30$.}
    \label{disc_comparison}
\end{figure}

The infinite system of equations Eqn.~(\ref{trantersystem}) is truncated and inverted to solve for the coefficients $a_m$. We find that keeping the first twenty terms is generally sufficient, so the computation is fast, and the flow field at the surface can be reconstructed from the solution via Eqn.~(\ref{hankelinversion}). Analytical formulae exist for $\Lambda_n$ and $M_{mn}$ in the case where $H$ is either vanishingly small or infinitely large. In the remaining cases, these integrals must be computed numerically, for instance with an asymptotic expansion that exploits the rapid decay of $\tanh kH$ to unity or a specialized computational package for oscillatory integrals like the IIPBF package developed by~\cite{Ratnanather_etal2014}. The results obtained using Tranter's method are in excellent agreement with those obtained by using the Green's function approach, as demonstrated in Fig.~\ref{disc_comparison}.


\subsection{Asymptotic solution when $H\to 0$}
\label{smallHlimit}

In the limit of vanishing subphase thickness, the substrate drag dominates. This can be seen by letting $H\to 0$ in Eqn.~(\ref{FTcond}), which reduces to the simple condition
\begin{equation}
\hat{\boldsymbol{f}} = \Gamma \hat{\boldsymbol{U}},
\end{equation}
where $\Gamma = \mu / H$ is the substrate drag coefficient. Inverting the Fourier transform yields the Brinkman equation
\begin{equation}
- \nabla P + \bar\eta\bnabla^2 \boldsymbol{U} = \Gamma \boldsymbol{U}
\label{brinkmaneqn}
\end{equation}
whose general solution in the axisymmetric case is of the form $V(r) = C K_1(r/\bar{\delta}) + D I_1(r/\bar{\delta})$, where $\bar{\delta}^2 = \bar\eta/\Gamma$, $I_1$ and $K_1$ are modified Bessel functions, and $C$ and $D$ are constants. In order to avoid a  blowup at the origin, $C=0$; the constant $D$ is then found by applying the stress boundary condition Eqn.~(\ref{robinbc}) at $r=R$, ultimately yielding the solution
\begin{equation}
V(r) = 2\Omega \bar{\delta} \frac{\eta_R I_1\left(r/\bar{\delta}\right) }{\eta I_2\left(R/\bar{\delta}\right) + \eta_R I_0\left(R/\bar{\delta}\right)}\chi( r < R).
\label{smallHsol}
\end{equation}
Note this function is discontinuous at $r=R$. Figure~\ref{tvimg} shows Eqn.~(\ref{smallHsol}) is the asymptotic limit of the solution to Eqn.~(\ref{dropleteqns}) as $H\to 0$. In this ``high friction case,'' the flow is largely confined to a boundary layer of width $\bar{\delta}$.

\section{An annular domain}
\label{sec:annulus}

We now take $\mathcal{D}$ to be the annulus with radii $0 < R_i(t) <R_o(t)$.
Unlike the disc case, there will be a radial component to the flow field in addition to a azimuthal one, and the odd viscosity will play a nontrivial role in this geometry. 

The divergence of the surface velocity field is assuredly nonzero since the interface moves inward; however, for $R_i<r<R_o$, the divergence is still zero by assumption. This condition restricts the radial flow to something of the form
\begin{equation}
U(r) = \frac{F}{r}\text{ , }R_i<r<R_o
\end{equation}
for some constant $F<0$, to be determined. The operator $\mathcal{L}$ annihilates $U$ on $R_i<r<R_o$, and the radial momentum equation simply reduces to
\begin{equation}
-\frac{\mathrm{d}P}{\mathrm{d}r} = f,
\label{radialpressureeqn}
\end{equation}
where 
\begin{equation}
f =\left. \frac{\partial u}{ \partial z} \right|_{z=0}
\end{equation}
 obeys Eqn.~(\ref{ftradial}) when Hankel transformed.
While Eqn.~(\ref{radialpressureeqn}) seems to imply the radial bulk dynamics are completely independent of the three monolayer viscosities in this geometry, we recall the definition of $P$ contains $\eta_O$.
If we take the surface pressure outside of the annulus to be zero, the radial and azimuthal stress boundary conditions for the annulus are found to be
\begin{equation}
\left.-P(r)-\frac{2\eta F}{r^2} + 2\eta_O \frac{V(r)}{r} \mp \frac{\gamma}{r}\right|_{r=R_i,R_o} = 0
\label{bcradial}
\end{equation}
\begin{equation}
\left.-\bar\eta V'(r)  +2\eta_R \Omega + (\eta-\eta_R) \frac{V(r)}{r}  + 2\eta_O \frac{F}{r^2} \right|_{r=R_i,R_o} = 0
\label{bctangential}
\end{equation}
where in the radial boundary conditions, the negative sign corresponds to $r=R_i$ and the positive sign corresponds to $r=R_o$. Note that when $\eta_O\neq0$, the radial and azimuthal flows are coupled through the boundary conditions.

\subsection{Solution via Green's function}

The solution to the annulus problem using the Green's function formulation mirrors that of the disc problem. We begin with the azimuthal velocity which is again decomposed as
\begin{equation}
V(r) = V_p(r) + V_h(r),
\end{equation}
where the definition of $V_p$ is slightly modified to account for the new limits of integration,
\begin{equation}
V_p(r) = \int_{R_i}^{R_o} \mathrm{d}r'~ G(r,r') g(r') = - \frac{1}{2\bar\eta r} \int_{R_i}^r \mathrm{d}r'~r'^2 g(r') - \frac{r}{2\bar\eta} \int_r^{R_o} \mathrm{d}r' ~g(r'),
\label{Vpdef2}
\end{equation}
and the definition of $V_h$ remains intact. For the annulus, the constants $A_h$ and $B_h$ satisfy
\begin{equation}
V_p'(R_i^+) + A_h -\frac{B_h}{R_i^2} - \frac{\eta-\eta_R}{\bar\eta R_i} \left[V_p(R_i) + A_hR_i +\frac{B_h}{R_i}\right]= \frac{2\eta_R\Omega}{\bar\eta}
\end{equation}
\begin{equation}
V_p'(R_o^-) + A_h -\frac{B_h}{R_o^2} - \frac{\eta-\eta_R}{\bar\eta R_o} \left[V_p(R_o) + A_hR_o +\frac{B_h}{R_o}\right]= \frac{2\eta_R\Omega}{\bar\eta} 
\end{equation}
where $R_i^+$ and $R_o^-$ are right and left limits, respectively. Solving for $A_h$ and $B_h$,
\begin{equation}
A_h = {\Omega} +\frac{ (\eta-\eta_R)[R_iV_p(R_i) - R_oV_p(R_o)] + \bar\eta [R_o^2 V_p'(R_o^-) - R_i^2 V_p'(R_i^+)]}{2\eta_R (R_i^2-R_o^2)}
\end{equation}
\begin{equation}
B_h = -\frac{2\eta_OF}{\eta} + \left[\frac{\eta-\eta_R}{2\eta} \left(\frac{V_p(R_i)}{R_i} - \frac{V_p(R_o)}{R_o}\right) + \frac{\bar\eta}{2\eta}[V_p'(R_o^-)-V_p'(R_i^+)]\right]\left(\frac{1}{R_o^2}-\frac{1}{R_i^2}\right)^{-1}.
\end{equation}
From Eqn.~(\ref{Vpdef2}), we obtain the identities
\begin{equation}
R_iV_p(R_i) - R_oV_p(R_o) = \frac{1}{2\bar\eta} \int_{R_i}^{R_o} \mathrm{d}r'~ (r'^2 -R_i^2  )g(r')
\end{equation}
\begin{equation}
R_o^2V'_p(R_o^-) - R_i^2V_p'(R_i^+) = \frac{1}{2\bar\eta} \int_{R_i}^{R_o} \mathrm{d}r'~ (r'^2 +R_i^2) g(r')
\end{equation}
\begin{equation}
\frac{V_p(R_i)}{R_i} - \frac{V_p(R_o)}{R_o} = \frac{1}{2\bar\eta} \int_{R_i}^{R_o} \mathrm{d}r'~ \left(-1 + \frac{r'^2}{R_o^2}\right) g(r')
\end{equation}
\begin{equation}
V_p'(R_o^-) - V_p'(R_i^+) = \frac{1}{2\bar\eta} \int_{R_i}^{R_o} \mathrm{d}r'~ \left(1 + \frac{r'^2}{R_o^2}  \right)g(r').
\end{equation}
Substituting everything into Eqn.~(\ref{Vsie}), we finally have
\begin{align}
&\int_{R_i}^{R_o} \mathrm{d}r'~ G(r,r') g(r') + r\left[{\Omega} +\frac{\eta}{2\eta_R\bar\eta(R_i^2-R_o^2)} \int_{R_i}^{R_o} \mathrm{d}r'~ r'^2 g(r') + \frac{R_i^2}{2\bar\eta (R_i^2-R_o^2)} \int_{R_i}^{R_o} \mathrm{d}r' ~g(r') \right]\nonumber\\
&+\frac{1}{r} \left[-\frac{2\eta_O F}{\eta} + \frac{\rho}{2\bar\eta R_o^2}\int_{R_i}^{R_o} \mathrm{d}r' ~r'^2 g(r') + \frac{\eta_R \rho}{2\bar\eta \eta} \int_{R_i}^{R_o} \mathrm{d}r' ~g(r')\right] = \frac{1}{\mu}\int_{R_i}^{R_o} dr'~r' L(r,r') g(r')
\end{align}
where $\rho = [(1/R_o)^2 -(1/R_i)^2]^{-1}$. As before, this integral equation can be inverted for $g$ as a function of $F$, which in turn gives $V$ in terms of $F$. Note that the definition of $L_d$ changes due to the new limits of integration:
\begin{align}
    L_d(r) &= \int_{R_i}^{R_o} \mathrm{d}r'~r'\bar{L}(r,r') \nonumber \\
    &= \int_0^\infty \mathrm{d}k~ J_1(kr) \int_{R_i}^{R_o}\mathrm{d}r'~r'J_1(kr') \nonumber \\
    &=-\frac{R_i^3}{6r^2}{}_3F_2\left[\{\frac{1}{2},\frac{3}{2},\frac{3}{2}\},\{2,\frac{5}{2}\},\frac{R_i^2}{r^2}\right] \nonumber \\
    &\quad + \frac{r}{32}\left(16\log \frac{4R_o}{r}-8-\frac{3r^2}{R_o^2}~{}_4F_3\left[\{1,1,\frac{3}{2},\frac{5}{2}\},\{2,2,3\},\frac{r^2}{R_o^2}\right]\right).
\end{align}
Also note that if $R_i$ and $F$ are allowed to approach zero, the integral equation reduces to that of the disc case,  Eqn.~(\ref{gie_disc}), with $R=R_o$.

\begin{figure}
\centering
\includegraphics[scale=0.16]{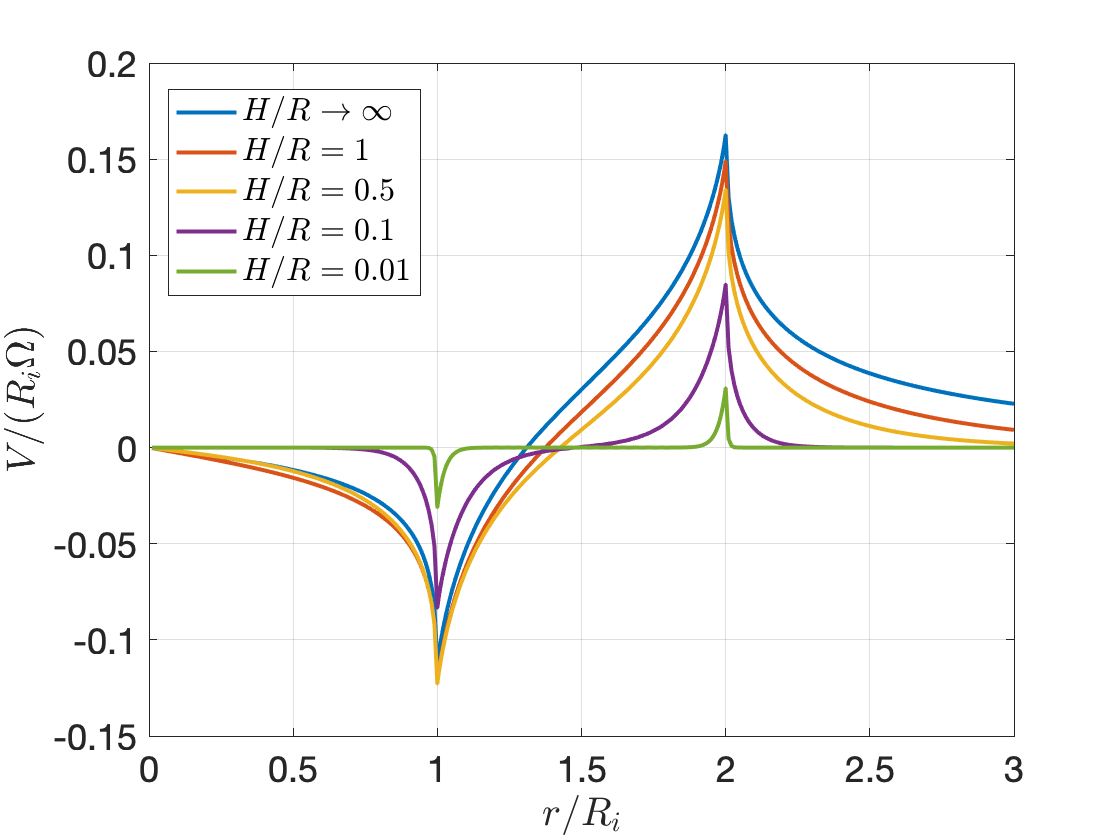}
\includegraphics[scale=0.16]{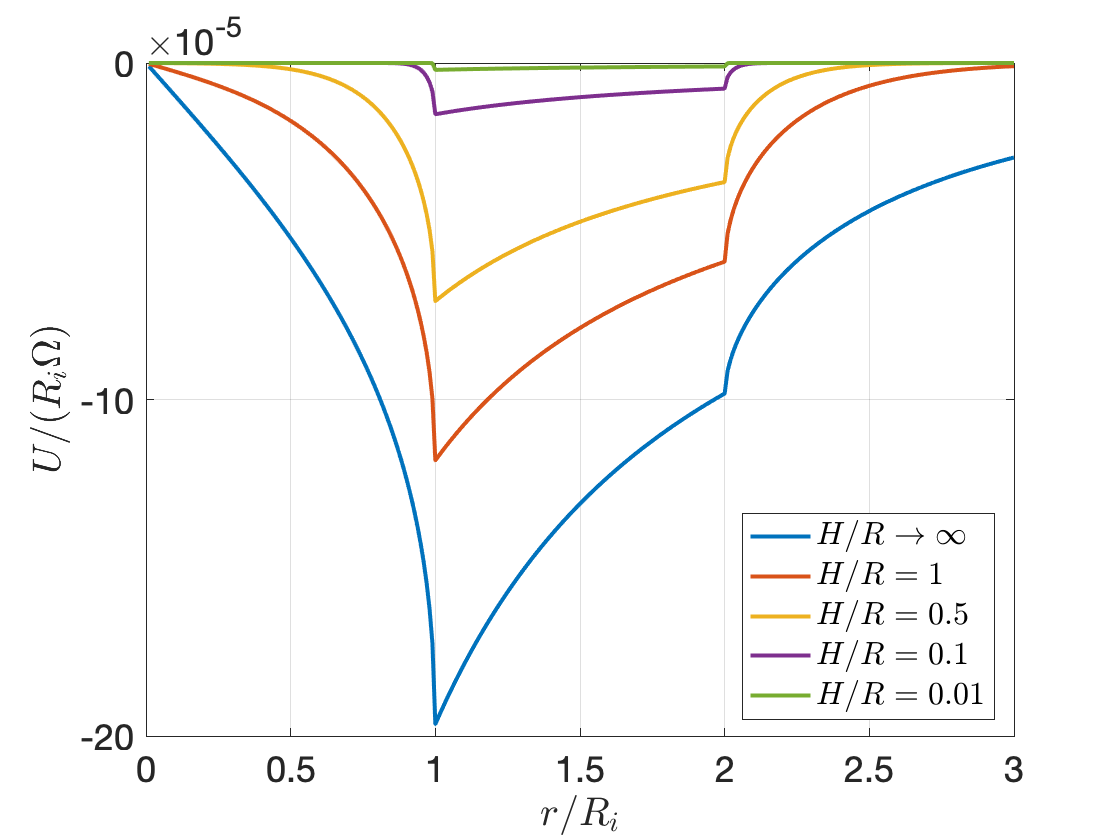}
\includegraphics[scale=0.16]{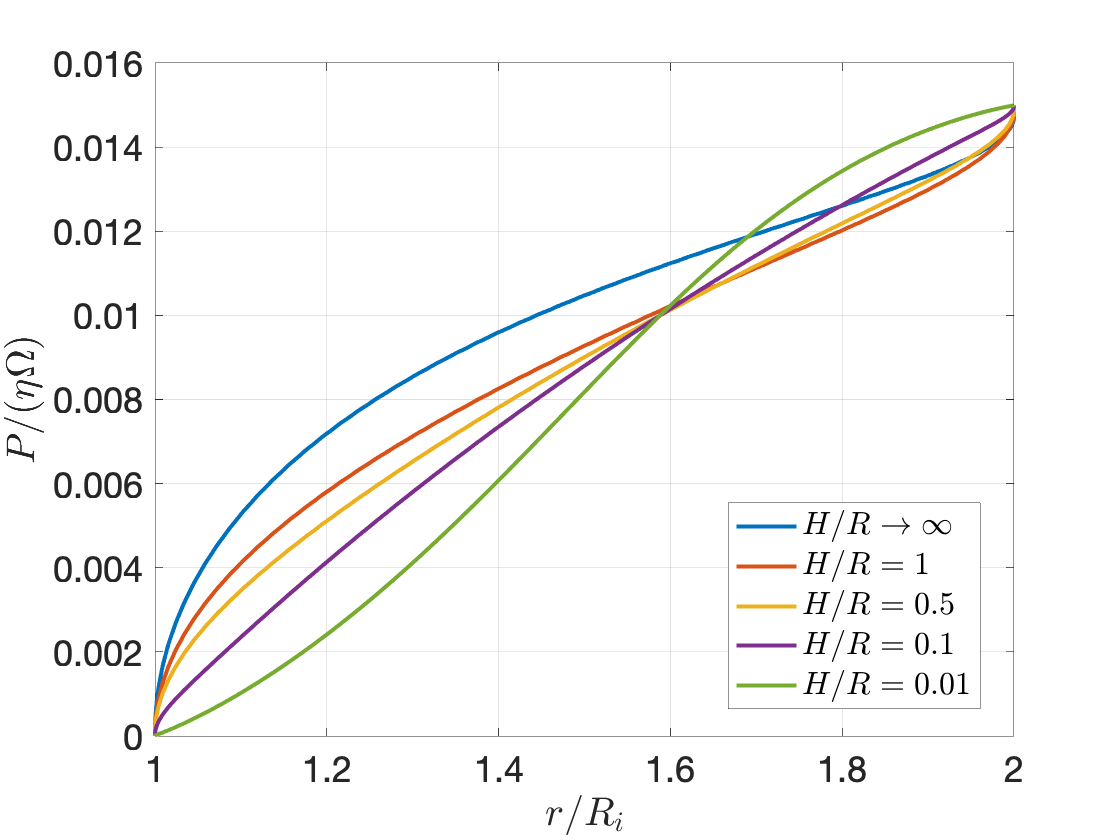}
\caption{The instantaneous surface azimuthal velocity $V$ (top left) and radial velocity $U$ (top right) as well as the monolayer pressure $P$ (bottom) due to an annular monolayer of radii $R_o/R_i = 2$ as subphase depth $H$ is varied. As $H\to 0$, boundary layers of width $\bar{\delta} = \sqrt{\bar\eta H/\mu}$ becomes visible in $V$. Parameters: $\eta_R/\eta =1.875 $, $\mu R/\eta = 30$, $\eta_O/\eta = 0$, $\tau_1/\tau_2=\gamma/(\eta R_{i}\Omega) = 0.01$. Since $\ell_{SD}\ll R$, $V$ scales like $\ell_{SD}/\tau_1 = \eta\Omega/\mu$ and $U$ scales like $\ell_{SD}/\tau_2=\gamma/(\mu R_i)$.}
\label{annulusvelocity}
\end{figure}

\begin{figure}
\centering
\includegraphics[scale=0.16]{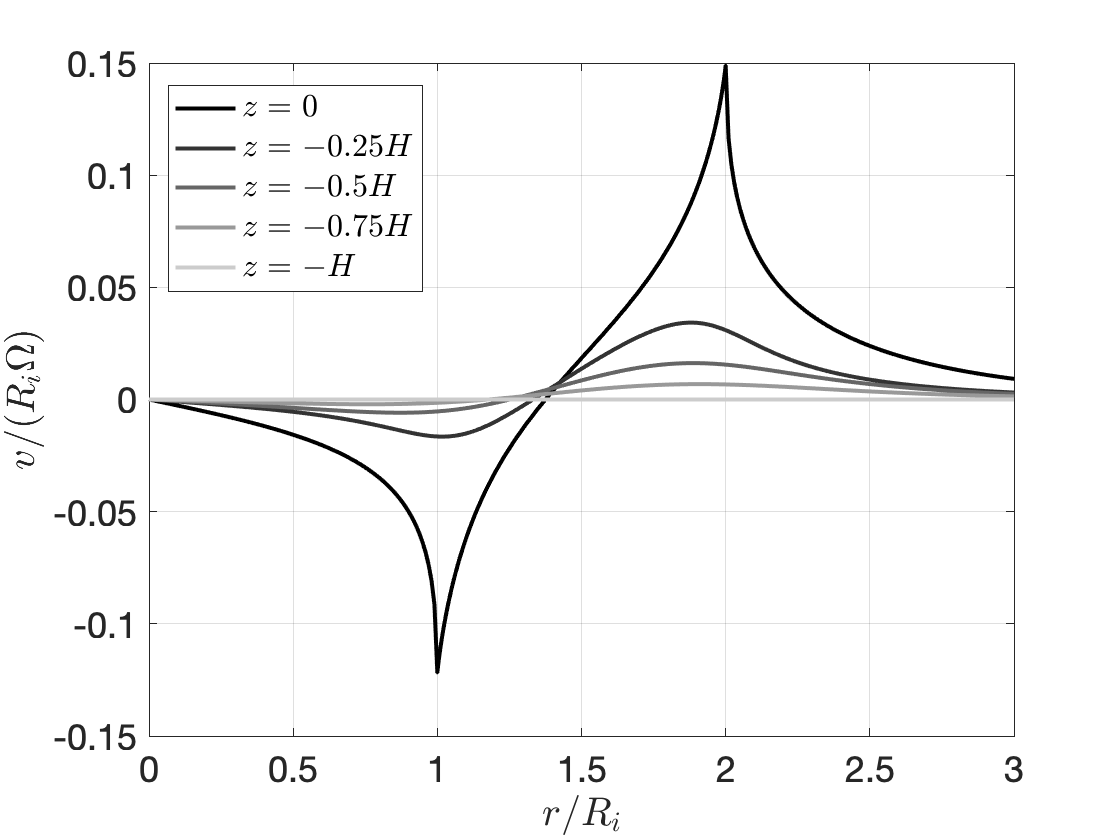}
\includegraphics[scale=0.16]{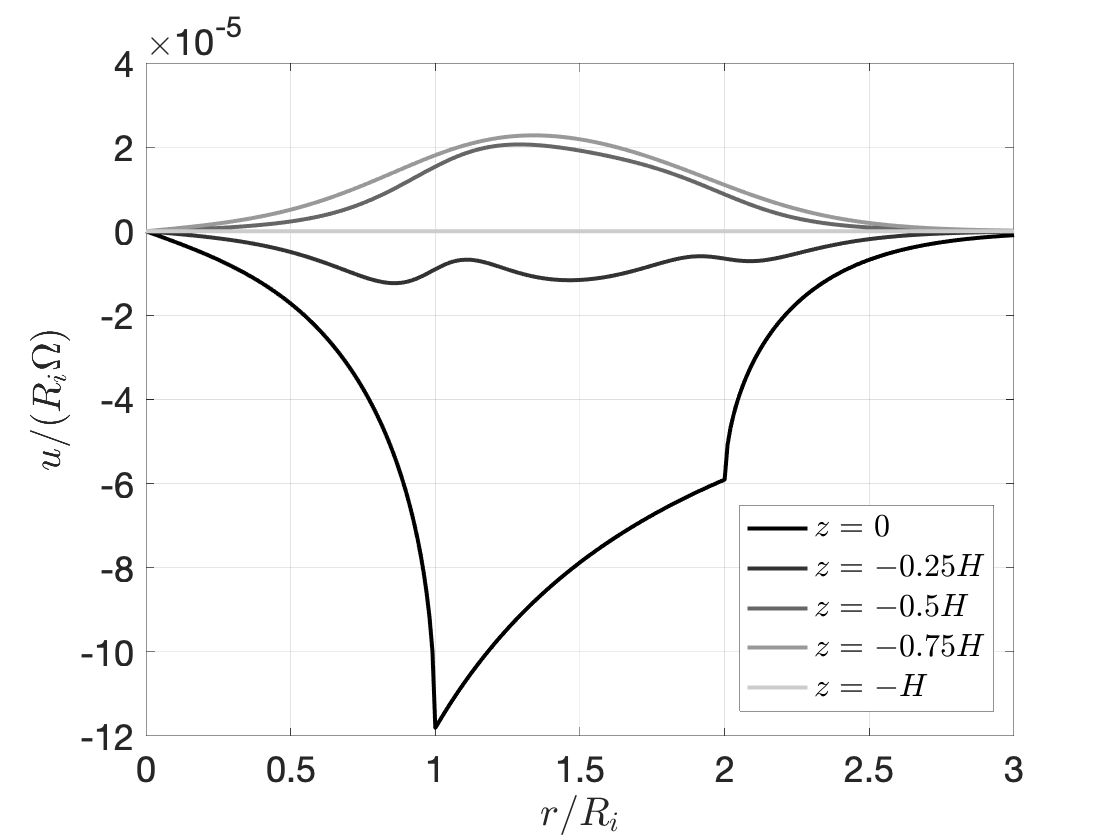}
\includegraphics[scale=0.16]{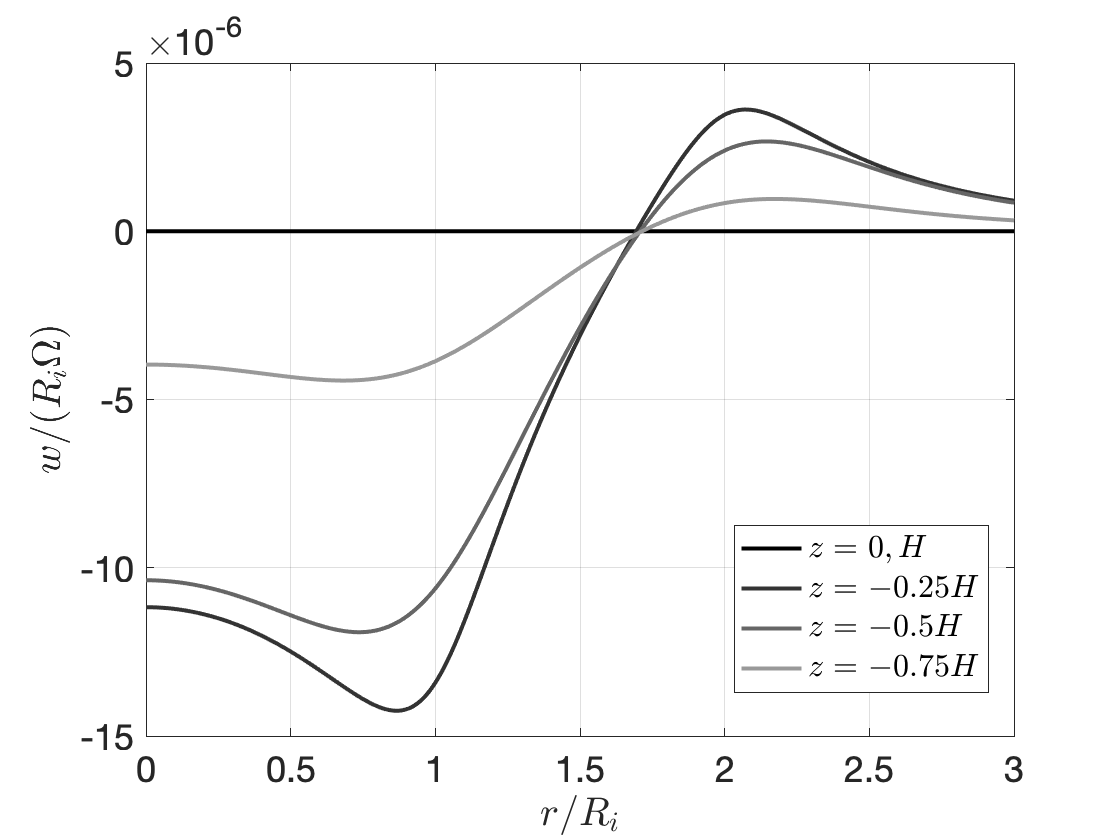}
\caption{The azimuthal component $v$ (top left), radial component $u$ (top right), and vertical component $w$ (bottom) of the velocity field $\boldsymbol{u}$ in the subphase at various depths for an annular monolayer of $R_o/R_i=2$ and $H=1$.  Parameters: $\eta_R/\eta =1.875 $, $\mu R/\eta = 30$, $\eta_O/\eta = 0$, $\gamma/(\eta R_{i}\Omega) = 0.01$. }
\label{annulussubphase}
\end{figure}

With the azimuthal velocity essentially solved, we turn to the radial velocity. Since $\mathcal{L}[U]$ vanishes in Eqn.~(\ref{Usie}), we are left with
\begin{equation}
    \frac{F}{r} = \frac{1}{\mu} \int_{R_i}^{R_o}\mathrm{d}r' ~r' M(r,r') f(r').
\end{equation}
By expanding $M(r,r')$ in the same way as $L(r,r')$ above, this integral equation can be numerically inverted to solve for $f/F$. Since $V$ is known (up to $F$), the constant $F$ can be determined from the fact that
\begin{equation}
    P(R_o^-) - P(R_i^+) = -\int_{R_i}^{R_o} \mathrm{d}r'~ f(r')
\end{equation}
along with Eqn.~(\ref{bcradial}). This completes the solution of the instantaneous surface flow field. Representative solutions for different values of $H$ are shown in Fig.~\ref{annulusvelocity}. The scaling arguments established in the disc case carry over: in the $\bar\beta \gg 1$ limit, the appropriate length scale for the velocity is $R$, while it is $\ell_{SD}$ in the small $\bar\beta$ limit. However, the radial and azimuthal components have different time scales.
For instance, in Fig.~\ref{annulusvelocity}, $U$ and $V$ exhibit a disparity in scale, with $U$ being approximately a hundred times smaller in magnitude than $V$. This can be traced back to Eqns.~(\ref{frenetbcrad}) and~(\ref{frenetbctan}), which show that in the $\eta_O=0$ case, the azimuthal motion originates from the rotational drive while the radial motion originates from the line tension. The corresponding time scales are $\tau_1 = \Omega^{-1}$ and $\tau_2=\eta R_i/\gamma$; for the curves plotted in Fig.~\ref{annulusvelocity}, the ratio of these time scales was taken to be $\tau_1/\tau_2 = \gamma/(\eta R_i \Omega) = 0.01$. Thus, for $\bar\beta \gg 1$, $V\sim R_i/\tau_1=R_i\Omega$ and $U\sim R_i/\tau_2=\gamma/\eta$, while for $\bar\beta \ll 1$, $V\sim \ell_{SD}/\tau_1= \eta\Omega/\mu$ and $U\sim \ell_{SD}/\tau_2 = \gamma/(\mu R_i)$.
Figure~\ref{annulussubphase} illustrates the subphase velocity field for the $H=R_i$ case calculated from Eqns.~(\ref{vbulk})-(\ref{wbulk}). 
Dynamics follow from advection of the domain boundary according to the kinematic boundary condition, which we describe in more detail in \textsection\ref{holeclosure}.

\subsection{Formulation as triple integral equations}


Similar to the disc case, we can convert the problem into two sets of triple integral equations. Keeping the definition of $a(k)$ from Eqn.~(\ref{adef}), the azimuthal set is
\begin{equation}
\int_0^\infty \mathrm{d}k ~k a(k) J_1(kr)=0\text{ , $0<r<R_i$}
\label{tangentialte1}
\end{equation}
\begin{equation}
\int_0^\infty \mathrm{d}k~ k \left(\frac{\bar\eta}{\mu} k \tanh kH + 1\right) a(k)J_1(kr)=0 \text{ , $R_i<r<R_o$}
\label{tangentialte2}
\end{equation}
 \begin{equation}
\int_0^\infty \mathrm{d}k ~k a(k) J_1(kr)=0\text{ , $r>R_o$}
\label{tangentialte3}
\end{equation}
and the radial set is
\begin{equation}
\int_0^\infty \mathrm{d}k~ k b(k) J_1(kr)=0\text{ , $0<r<R_i$}
\label{radialte1}
\end{equation}
\begin{equation}
 \int_0^\infty \mathrm{d}k ~b(k)[A(kH) + B(kH)]^{-1}J_1(kr)=\frac{\mu F}{r} \text{ , $R_i<r<R_o$}
 \label{radialte2}
\end{equation}
 \begin{equation}
\int_0^\infty \mathrm{d}k ~k b(k) J_1(kr)=0\text{ , $r>R_o$}
\label{radialte3}
\end{equation}
where we have defined
\begin{equation}
b(k) = \int_0^\infty \mathrm{d}r~ r f(r) J_1(kr)
\end{equation}
and used Eqn.~(\ref{ftradial}).

Difficulties similar to the dual integral equations of the previous section plague the azimuthal triple integral equations. In particular, Eqns.~(\ref{tangentialte1}-\ref{tangentialte3}) again erroneously appear to be homogeneous because the problem as stated above is not closed. Unfortunately, no convenient basis analogous to Tranter's for the disc appears to resolve the problem; attempting to use the Tranter basis as before will lead to a result that is discontinuous at $r=R_i$, as the basis is ``unaware'' of the divergence there. One possible workaround is to Fourier transform the equations to include boundary conditions and discretize the equations directly on the interval $(0,\infty)$ to yield a large linear system of equations. However, this method converges very slowly and performing it repeatedly to time step the kinematic boundary condition is not practical. Thus, we content ourselves with solving the equations using the aforementioned singular integral formulation.

On the other hand, several methods exist for solving the radial triple integral equations.
One approach due to~\cite{Cooke1965} involving Erd\'{e}lyi-Kober operators (generalized fractional derivatives) can be used to solve the radial equations, up to the unknown constant $F$. In the simpler case of $H\to\infty$, a different method also devised by~\cite{Cooke1963} involving rewriting the equation as a composition of Abel transforms (see~\cite{Noble1958}) can be used as well. These methods reduce the set of triple integral equations to a single Fredholm integral equation, which must be numerically solved, ultimately making them more work than the solution we have presented.

It is notable that the radial set of triple integral equations has a simple exact solution in the limit $H,R_o\to \infty$, as found by \cite{Alexander_etal2006}. In this limit,
\begin{equation}
    b(k)= \mu F\frac{\sin kR_i}{k R_i}
    \label{bexact}
\end{equation}
which, via the Hankel inversion theorem, leads to (cf.~\cite{GradshteynRyzhik2007} Formula (6.693.1) and differentiate under the integral)
\begin{equation}
    -\frac{\mathrm{d}P}{\mathrm{d} r} =\frac{\mu F}{r\sqrt{r^2-R_i^2}}\chi({r>R_i})
\end{equation}
so that the pressure (with constant of integration zero) is
\begin{equation}
    P =-\mu F \frac{\cos^{-1}(R_i/r)}{R_i}
\end{equation}
for $r>R_i$, and
\begin{equation}
    U = \frac{F}{r}\left(1-\sqrt{1-\frac{r^2}{R_i^2}}\right)
\end{equation}
for $0<r<R_i$. For $r>R_i$, $U=F/r$ as previously. 


\subsection{Asymptotic solution when $H\to 0$}
\label{smallHsol_ann}

\begin{figure}
    \centering
    \includegraphics[scale=0.2]{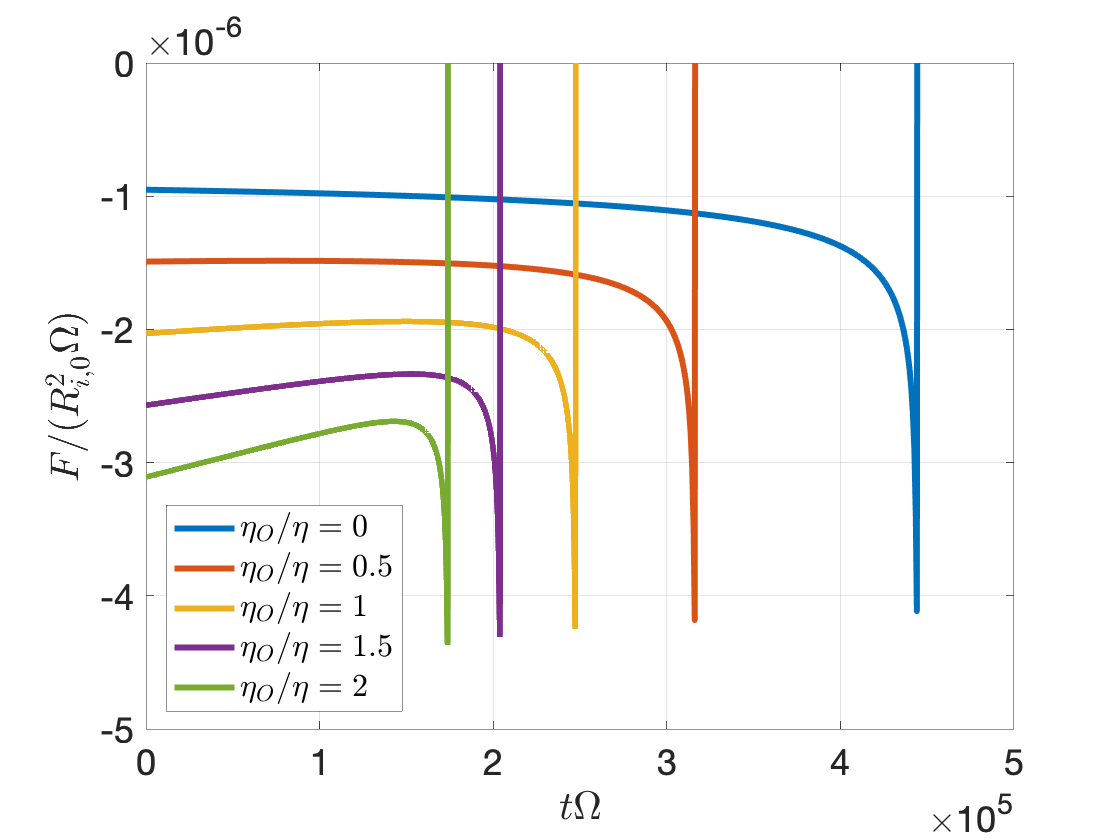}
    \caption{The parameter $F$ as a function of time for various odd viscosities in the high friction case. The initial slope depends on $\eta_O$; if $\eta_O/\eta$ is sufficiently large $F$ has a local maximum. As the hole closes, $F$ reaches a local minimum and then rapidly approaches zero in a small shear viscosity dominated regime so that $F(t)$ appears nearly vertical. Initially, $R_{o,0}/R_{i,0} = 5$. Curves from right to left: $\eta_O/\eta = 0,0.5,1,1.5,2$. Parameters: $\eta_R/\eta = 1.857\times 10^{-2}$, $\Gamma R_{i,0}^2/\eta  = 4.074\times 10^{2}$, $\gamma/(\eta R_{i,0} \Omega) = 5.214\times 10^{-4}$. }
    \label{highfrictionF}
\end{figure}

As with the disc-shaped domain, the small $H$ case reduces to a Brinkman equation; we will impose $\boldsymbol{U} = \boldsymbol{0}$ on $\mathcal{D}^C$ so that $\bnabla \bcdot \boldsymbol{U} = 0$ everywhere and the drag is isotropic. The general solution in the axisymmetric case is
\begin{equation}
U(r) = \frac{F}{r},
\end{equation}
\begin{equation}
V(r) = CK_1\left(\frac{r}{\bar{\delta}}\right) + DI_1\left(\frac{r}{\bar{\delta}}\right),
\end{equation}
\begin{equation}
P(r) = -\Gamma F \log r + G.
\end{equation}
The four unknowns $C,D,F$, and $G$ are found by substituting these expressions into the four boundary conditions Eqns.~(\ref{bcradial})-(\ref{bctangential}). Exact but cumbersome expressions for these constants can be found; in the interest of brevity, we omit them. However, in the simple case of zero odd viscosity, the constant $F$ is found to be
\begin{equation}
    F = -\frac{\gamma \left[(1/R_i) + (1/R_o)\right]}{\Gamma \log (R_o/R_i) +2\eta [(1/R_i^{2}) -(1/R_o^{2})]},
\end{equation}
which, through the kinematic boundary condition, determines the size of the cavity. Note that $R_o$ is related to $R_i$ through the monolayer incompressibility constraint: $A = \pi (R_o^2 -R_i^2)$. We remark that as in the disc case, boundary layers of width $\bar{\delta} = \sqrt{\bar\eta/\Gamma}$ are visible in the azimuthal velocity field in this limit (Fig.~\ref{annulusvelocity}). If $\bar{\delta}$ is sufficiently smaller than $R_i$, then we may employ the boundary layer approximations
$V'(R_i) \sim -V(R_i)/\bar{\delta}$ and $V'(R_o)\sim V(R_o)/\bar{\delta}$ to obtain
\begin{equation}
    \frac{V(R_i)}{R_i}\sim \frac{-2\eta_R \Omega - (2\eta_o F/R_i^2)}{(\bar\eta R_i/\bar{\delta}) + (\eta-\eta_R)}
\end{equation}
and similarly for $V(R_o)$. Inserting these into the radial boundary conditions gives the correction due to odd viscosity:
\begin{equation}
    F \sim - \frac{\gamma \left[(1/R_i) + (1/R_o)\right] +4\eta_O \eta_R \Omega q_O}{\Gamma \log (R_o/R_i)  + 2\eta [(1/R_i^2)-(1/R_o^2)] - 4\eta_O^2q_O}
\end{equation}
where
\begin{equation}
    q_O = \frac{1}{(-\bar\eta R_o/\bar{\delta}) + \eta -\eta_R} -\frac{1}{(\bar\eta R_i/\bar{\delta})  +\eta-\eta_R}.
\end{equation}
Figure~\ref{highfrictionF} shows the parameter $F$ as a function of time as $\eta_O$ is varied. The impact of this parameter on the closing of the cavity is discussed in the next section.

\subsection{Hole closure dynamics \& the effect of odd viscosity}
\label{holeclosure}

For the disc-shaped domain, the absence of a radial velocity means that the boundary never moves. The odd viscous stresses are in the radial direction but are offset by the pressure and hence have no effect on the domain shape or flow field.

For the annular domain, the kinematic boundary condition states that the radii of the circular boundaries will change according to the local radial surface velocity:
\begin{equation}
\frac{\mathrm{d}R_i}{\mathrm{d}t} = U(R_i) =\frac{F}{R_i}
\label{kbc}
\end{equation}
and similarly for the outer radius. Equivalently, the outer radius can be found by applying the constraint that the monolayer area $\upi(R_o^2 -R_i^2)$ is constant. 
Eqn.~(\ref{kbc}) can be numerically integrated to find $R_i(t)$, up until $R_i=0$, at which point the circular disc case is recovered. At each time step, we must solve for $F$ by solving for the flow field using the procedure outlined in the previous section.

We first obtain some simple limits for the high friction case. Following the discussion in Appendix B, the Saffman-Delbr\"uck length for the high friction case is $\bar\delta$. For $\bar\delta \gg R_i$, the expected radial velocity scale is $R_i/\tau_2 = \gamma/\eta$, while for $\bar{\delta} \ll R_i$, it is $\bar{\delta}/\tau_2 = \gamma \bar{\delta}/(\eta R_i)$. In the initial phase of the experiment, the cavity radius is large compared to $\bar\delta$, so $F\sim \gamma \bar{\delta}/\eta$. However, as the hole is just about to close, $F\sim \gamma R_i/\eta$. These two regimes are illustrated in Fig.~\ref{highfrictionF}, where $F$ is initially nearly constant in the $\eta_O=0$ case and transitions to a linear regime with large slope when $R_i$ becomes comparable to $\bar\delta$.

\begin{figure}
\centering
\includegraphics[scale=0.165]{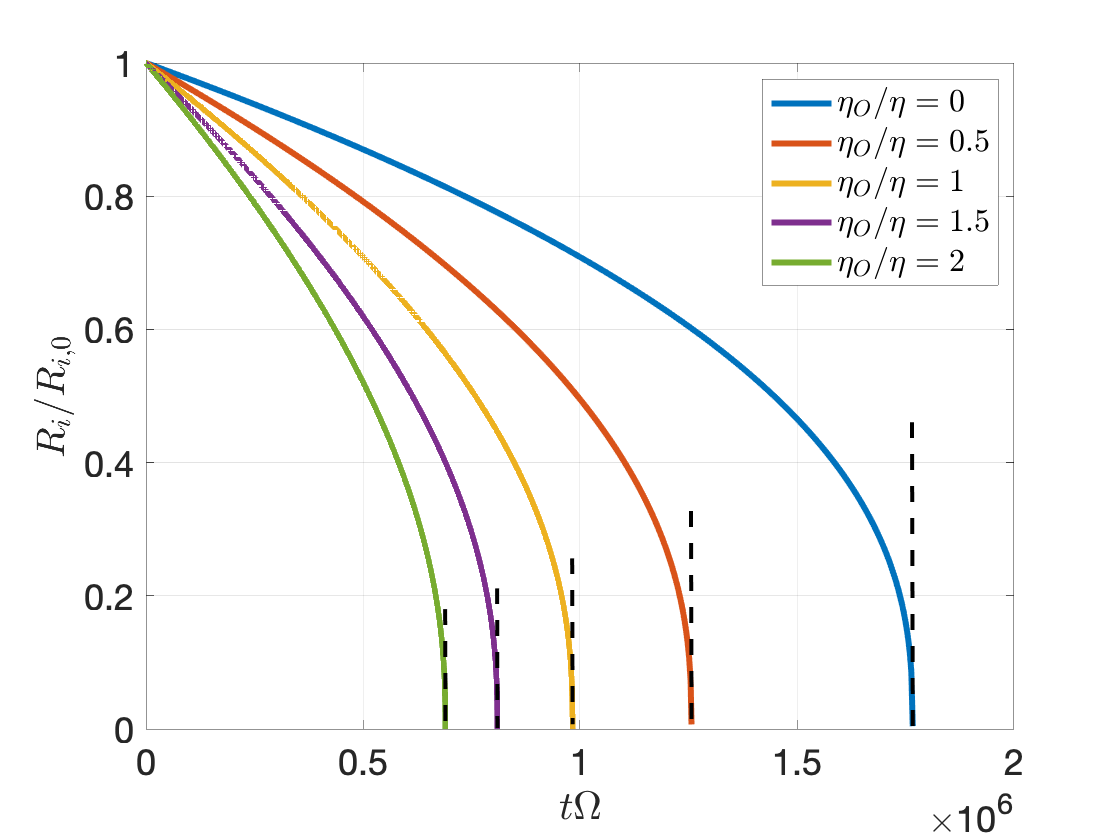}
\includegraphics[scale=0.165]{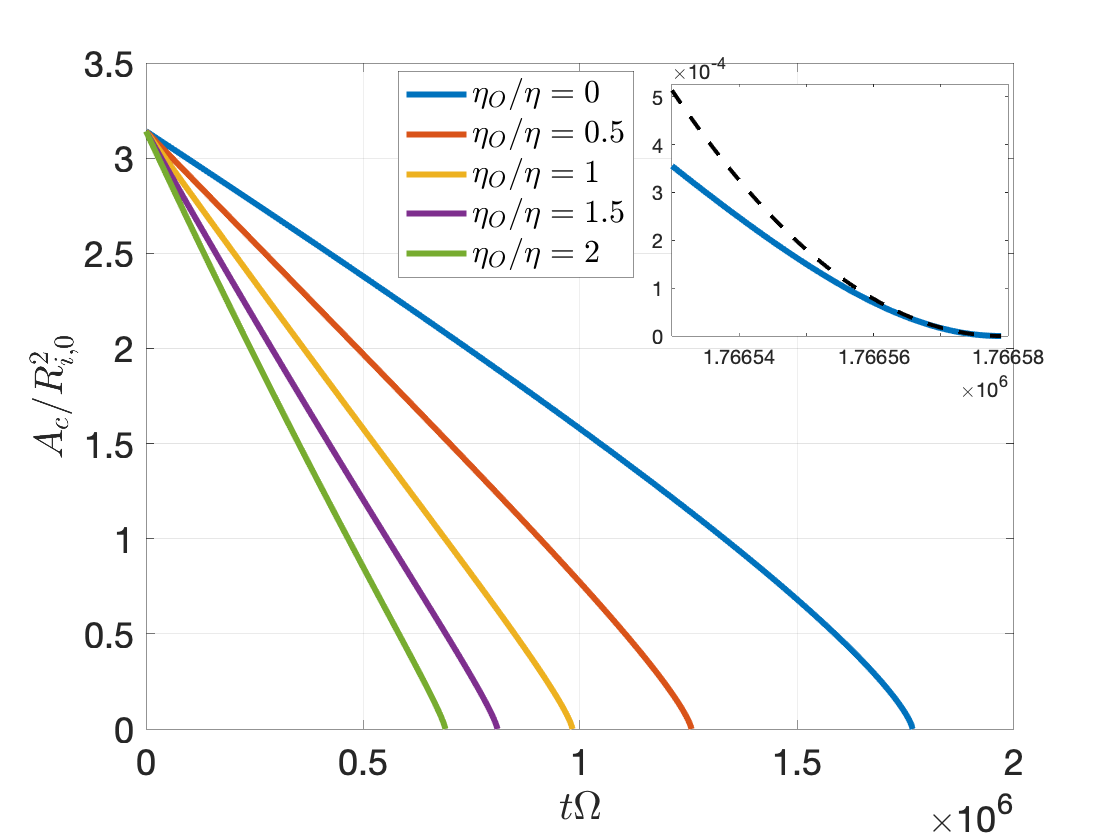}
\caption{(Left) Cavity radius as a function of time for various odd viscosity values in the high friction case. Dashed lines indicate the small hole asymptotic solution (note these lines are not vertical but have a large gradient). Initially, $R_{o,0}/R_{i,0} = 5$. Curves from right to left: $\eta_O/\eta = 0,0.5,1,1.5,2$. Parameters: $\eta_R/\eta = 1.857\times 10^{-2}$, $\Gamma/(\eta R_{i,0}^2) = 4.074\times 10^{2}$, $\gamma/(\eta R_{i,0} \Omega) = 5.214\times 10^{-4}$. (Right) Corresponding area of cavity $A_C$ as a function of time using the same parameters and color scheme. Increasing odd viscosity changes the apparent concavity of the cavity area vs. time curve. Inset: Larger version of the $\eta_O=0$ curve near $t=t^*$, where the hole size is comparable to the penetration depth $\bar\delta$. In this limit, the closing is dominated by shear viscosity and the area decreases quadratically.} 
\label{highfrictionfigs}
\end{figure}

An asymptotic analysis of the small $R_i$ limit reveals the closure time is finite. The argument proceeds as follows: 
for simplicity, assume that $H\to\infty$ and $V(R_i)\to 0$. If $R_i \ll R_o$, we can use the exact solution Eqn.~(\ref{bexact}) to calculate the pressure difference $P(R_o) - P(R_i) = -\mu \pi F/(2R_i)$, which is negligible compared to the shear viscous stress which scales like $F/R_i^{2}$. The boundary condition Eqn.~(\ref{bcradial}) shows that this stress must be balanced by the line tension, from which we find $F \sim -\gamma R_i/(2\eta)$. The kinematic boundary condition Eqn.~(\ref{kbc}) then implies $\mathrm{d}R_i/\mathrm{d}t$ is constant in this limit so that there is no blowup. Prior to this regime, the cavity area decreases at a nearly constant rate, and is well approximated by $A_C = \pi R_{i,0}^2 + 2\pi t F_0$, where $R_{i,0}$ is the initial cavity radius and $F_0$ is the value of $F$ at time $t=0$.
Figure~\ref{highfrictionfigs} illustrates these different regimes and compares the cavity radius and area as functions of time for different odd viscosities. Note that in the high friction case, the $t\to t^*$ shear viscosity-dominated regime occurs at a cavity radius comparable to $\bar\delta$, which is itself comparable to the particle size, and is thus not expected to be experimentally detectable. 

\begin{figure}
    \centering
    \includegraphics[scale=0.165]{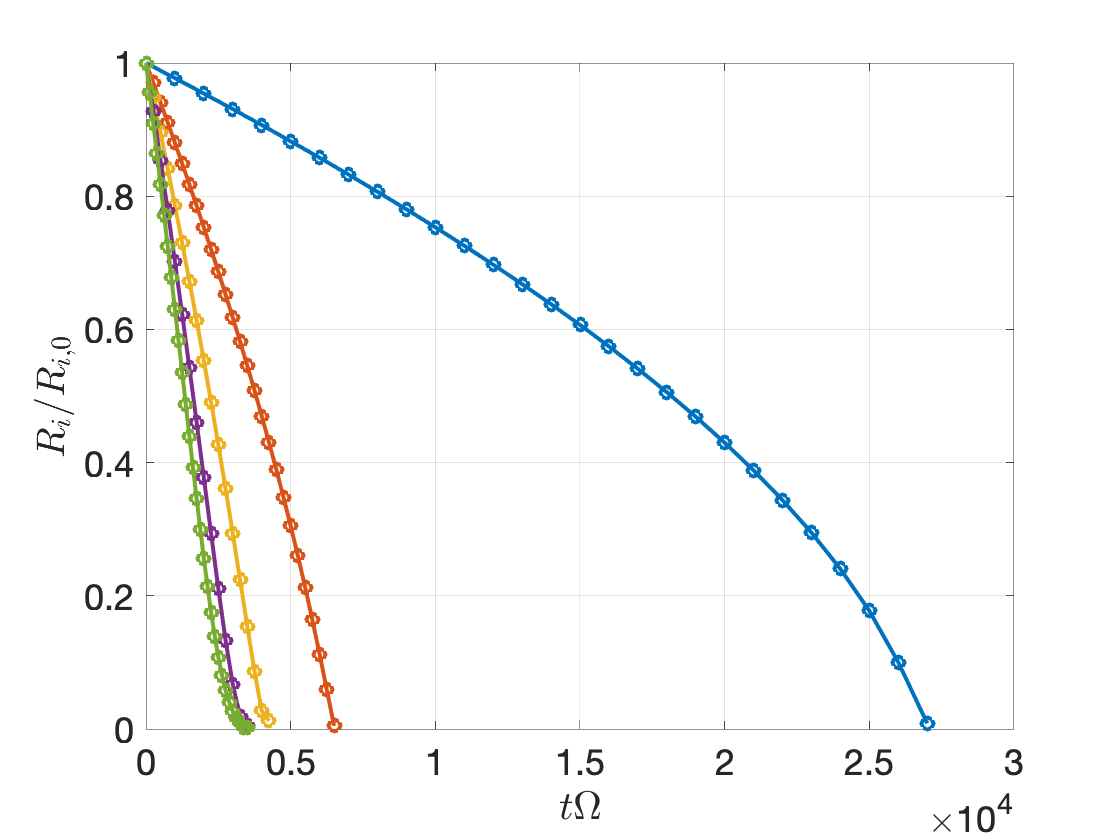}
    \includegraphics[scale=0.165]{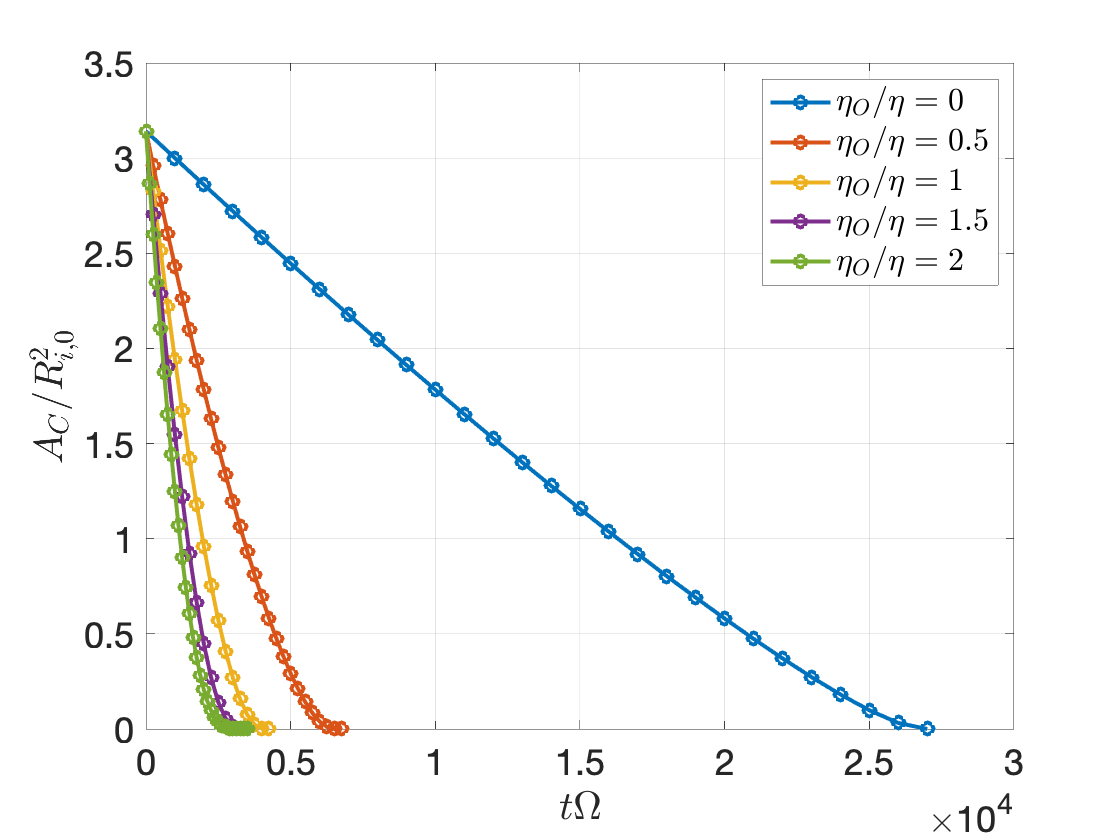}
    \caption{Cavity radius (left) and area (right) as functions of time for the low friction $H\to\infty$ case. Increasing the odd viscosity decreases the time needed to close the hole. The quadratic behavior of the cavity area in the shear viscosity-dominated regime is visible at larger radii compared to the high friction case. Parameters: $R_{o,0}/R_{i,0} =  5$, $\eta_R/\eta = 1.857\times 10^{-2}$, $\gamma/(\eta R_{i,0}\Omega) = 5.214\times 10^{-4}$, $\mu R_{i,0}/\eta = 8.79$. In the figure on the right, the transition from linear to quadratic behavior takes place when $R_i\sim \ell_{SD}=0.11R_{i,0}$.}
    \label{lowfrictionfigs}
\end{figure}

The most noticeable effect of odd viscosity is that it decreases the time it takes for the hole to close. If $\eta_O = 0$, the closure time is independent of $\Omega$ and $\eta_R$. On the other hand, a nonzero odd viscosity couples the azimuthal and radial velocities via the stress boundary conditions. 
At both the inner and outer boundaries, the line tension forces point radially inward toward the origin. The odd stress is oriented inward at the outer boundary but outward at the inner boundary; naively, one may think this causes the hole to close slower. However, this argument does not account for the pressure. As a consequence of domain incompressibility, the rate of change of the inner radius must be larger than that of the outer radius:
\begin{equation}
    \frac{\mathrm{d} R_i}{\mathrm{d} t } = \frac{R_o}{R_i}  \frac{\mathrm{d}R_o}{\mathrm{d}t}.
\end{equation}
This restriction implies that if the outer boundary is moving in faster with nonzero $\eta_O$, so too must the inner boundary. The odd viscosity also changes the concavity of the area vs. time curve. If $\eta_O\ll \eta$, the curve is concave (except for the shear viscosity-dominated region when the hole is very small, where it is always convex) but if $\eta_O$ is sufficiently large, it experiences regions of convexity as well. This behavior is not observed in the low friction case (Figure~\ref{lowfrictionfigs}).

\begin{figure}
    \centering
    \includegraphics[scale=0.165]{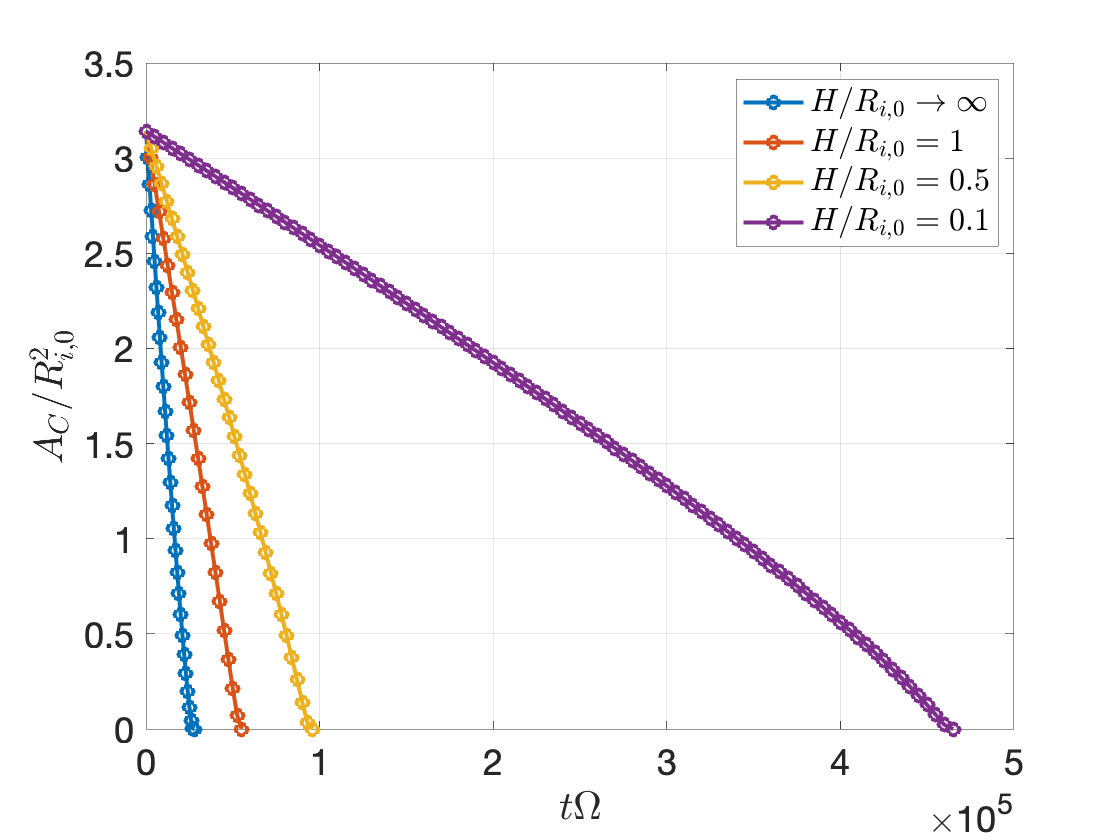}
    \includegraphics[scale=0.165]{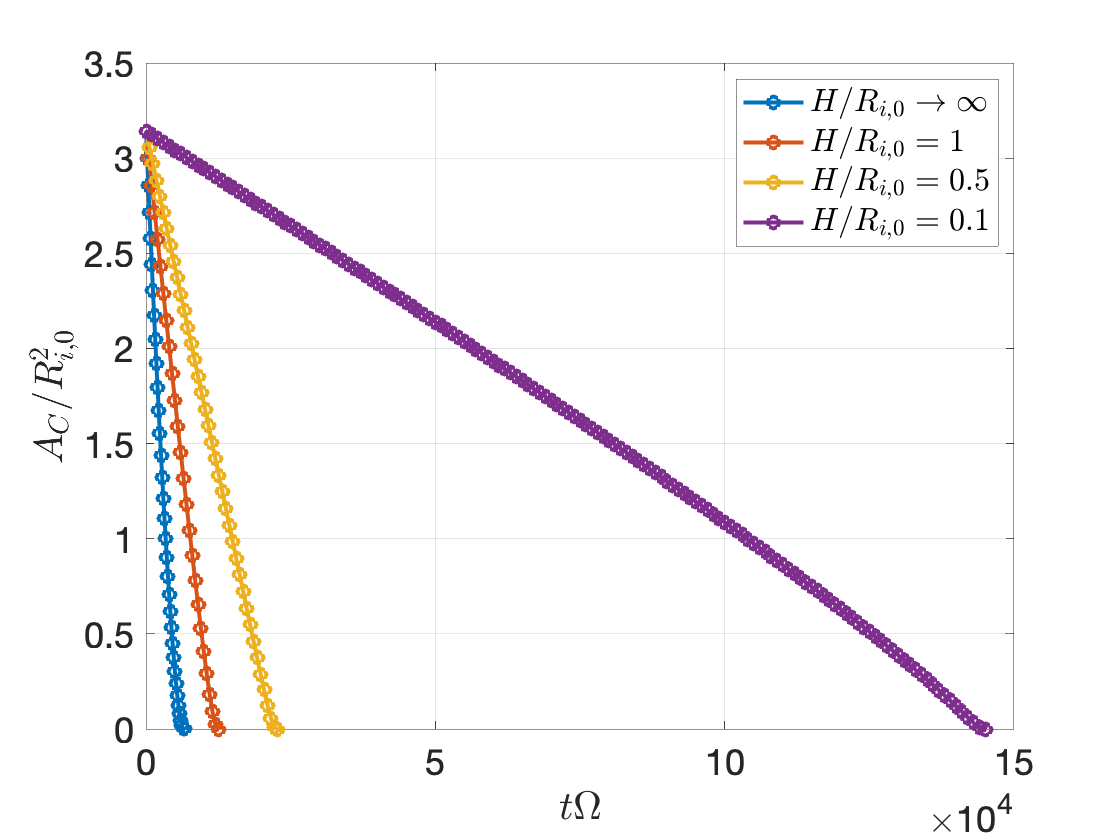}
    \caption{Increasing substrate height drastically decreases the time it takes for the hole to close. (Left) Parameters: $R_{o,0}/R_{i,0} =  5$,  $\eta_R/\eta = 1.857\times 10^{-2}$, $\eta_O/\eta = 0$, $\mu R_{i,0}/\eta  = 8.79$, $\gamma/(\eta R_{i,0}\Omega) = 5.214\times 10^{-4}$. (Right) Same parameters except $\eta_O/\eta = 0.5$.}
    \label{annulus_vary_H}
\end{figure}

For the low friction case, we find the same general trends. For zero odd viscosity, infinite substrate depth, and asymptotically large outer radius $R_o/R_i\to \infty$, a complete analytical description is possible using Eqn.~(\ref{bexact}). The cavity area as a function of time in this case is given by
\begin{equation}
    A_C=\frac{4}{\pi \mu^2} \left[2\eta -\sqrt{4\eta^2+\pi \gamma \mu (t-t^*)}\right]^2,
\end{equation}
where $t^* = (2\eta R_{i,0}+\pi\mu R_{i,0}^2/4)/\gamma$ is the closing time of the cavity written in terms of the initial radius $R_{i,0}$. When $R_i$ is large compared to $\ell_{SD}$, the area changes linearly in time with a constant $2\pi F\sim -4\gamma/\mu$. In the limit where $R_i\ll \ell_{SD}$, we find that $F\sim -\gamma R_i/(2\eta)$ just as in the high friction case (in fact, this limit is independent of $H$). Figure~\ref{lowfrictionfigs} shows the analogous radius and area curves as a function of time for the low friction case. The small radius viscosity-dominated regime is more clearly visible. The reader is referred to~\cite{JiaShelley2022} for details about this analytically tractable case.




\subsection{Conclusion and future work}

We have developed a formulation for the dynamics of an active, chiral surface phase coupled to a passive fluid underneath. We showed how to formulate the problem as calculating the surface velocity, given the surface stress, using a Green's function; this formulation is highly general and could be used to model other types of active (or passive) surface phases, or to study dynamics in more complicated, multi-connected domains with little to no modification. Using analytical and numerical methods, we proceeded to calculate the velocity fields for a disc-shaped and an annular monolayer.
For the case of a disc-shaped monolayer, a modification of Tranter's method allowed for a semi-analytical description and efficient numerical solution. For the case of an annulus, we thoroughly explored the effects of odd viscosity on the closing of a two-dimensional circular cavity. Our main results include an decrease in the cavity closure time in the presence of odd viscosity and a change in concavity of the cavity area vs. time curve as $\eta_O$ is increased in the high friction case. These results may provide another way to experimentally estimate the odd viscosity coefficient.

Ongoing work is focused in several different directions. Firstly, a boundary integral formulation for the high friction case to handle nonaxisymmetric shapes is under development. A full numerical formulation of the low friction case, much less its linear stability theory, is particularly challenging. Great care is needed to accommodate the divergent surface stresses, which are a fundamental part of the basic model. As a basic problem in applied mathematics, it would be interesting to find an analogue to the countable Tranter basis for the annular case and in that way develop a near analytical solution for its dynamics.

{\bf Acknowledgements:}
WTMI acknowledges support from the National Science Foundation under awards DMR-2011854 (University of Chicago MRSEC) and DMR-1905974. MJS acknowledges support by the National Science Foundation under awards DMR-1420073 (NYU MRSEC) and DMR-2004469. 
The authors are grateful to Florencio Balboa-Usabiaga, Ephraim Bililign, and Yehuda Ganan for helpful discussions.

\appendix
\section{The physical system and experimental parameter values}

The active chiral fluid we consider is a monolayer composed of thousands to millions of hematite particles, each roughly 1.6 microns in size and equipped with a magnetic dipole moment. The colloids are suspended in water and sedimented onto either a glass slide or an air/water interface; we refer to the former as the ``high friction case'' and the latter as the ``low friction case.'' Note that the particles are denser than water so that in the low friction case, the monolayer is found at the bottom of the water ``subphase,'' which is a top-down reflection of what is depicted in the schematic in Figure~\ref{fig:expcavity}. For the monolayer sizes considered here ($R\lesssim 500$ microns), the interface is well-approximated by an infinite plane, so for convenience, we may take the reflected configuration as our model without affecting any of our results. The depth of the water subphase, $H$, is typically comparable to $R$ in low friction experiments.

Under the application of an external rotating magnetic field, the particles spin; for frequencies in the range of roughly $\Omega =$ 2 to 12 Hz, the particles' rotational inertia is negligible so that the dipole moments are effectively always aligned with the external magnetic field. Since the average magnetic interaction is attractive, the system experiences effective surface and line tensions that form a cohesive two-dimensional incompressible fluid. \cite{Soni_etal2019} showed experimental examples of fluidic behavior and put forth a descriptive zero Reynolds number hydrodynamic theory accounting for three kinds of bulk viscous interparticle stresses: a shear viscous stress arising from  attractions between neighboring dipoles, a rotational stress arising from rotor-rotor friction, and an odd stress possibly arising from the collisions of rotating particles. 


Rheological tests by~\cite{Soni_etal2019} suggest that the shear viscosity $\eta$ of the colloidal fluid is around fifty times greater than its rotational viscosity $\eta_R$, while fitting the dispersion relation of low friction edge waves suggests that the odd viscosity $\eta_O$ is comparable in magnitude to the shear viscosity: $\eta = 4.9\pm 0.2 \times 10^{-8}$ Pa m s, $\eta_R = 9.1\pm 0.1 \times 10^{-10}$ Pa m s, and $\eta_O = 1.5\pm 0.1 \times 10^{-8}$ Pa m s. In the high friction case, these stresses are balanced against an external substrate friction $\Gamma = 2.49\pm 0.03 \times 10^3$ Pa s m$^{-1}$ that is generally found to be isotropic and proportional to the monolayer velocity. In the more complicated low friction case, the external forcing comes from the shear stress due to the motion of the fluid subphase with viscosity $\mu$, which is intimately coupled to that of the monolayer. 
At the boundary, the internal stresses are balanced by an edge tension $\gamma = 2.3\pm0.2 \times 10^{-13}$ N. The theoretical analysis in~\cite{Soni_etal2019} is restricted to the simpler high friction case of a monolayer situated on glass substrate; here, we will focus on the more general low friction case of a fluid subphase.

\section{Nondimensional groups}

It is instructive to consider the dimensionless versions of the monolayer momentum equation and associated boundary conditions. In the case of a monolayer with length scale $R$, we take the normal velocity scale to be $\bar{U}=\gamma/(\mu R)$, the tangential velocity scale to be $\bar{V} = R\Omega$, and the pressure scale to be $\mu \bar{U} =\gamma/R$. We begin by defining the dimensionless quantity $\alpha =\bar{U}/\bar{V}$. Temporarily identifying dimensionless quantities with their dimensional counterparts, the momentum equation for the monolayer in arbitrary coordinates that are consistent with the Frenet frame at the boundary becomes
\begin{equation}
-\bnabla P + \frac{\eta + \eta_R}{\mu R}\Delta (U, \alpha^{-1} V) = \left.{\mu} \frac{\partial \boldsymbol{u}(u,\alpha^{-1} v)}{\partial z} \right|_{z=0}
\end{equation}
which reveals the ratios of two types of Saffman-Delbr\"uck length to the monolayer size as two dimensionless parameters,
\begin{equation}
    \beta_S = \frac{\eta/\mu }{ R}
\end{equation}
\begin{equation}
    \beta_R = \frac{\eta_R/\mu}{ R}.
\end{equation}
Rescaling the stress boundary conditions in the same manner yields
\begin{equation}
   \left.  -P + 2\beta_S \left(\alpha^{-1} T_s -\kappa N\right) + 2\beta_O(N_s+\alpha^{-1}\kappa T) \right|_{\partial\mathcal{D}}= \kappa|_{\partial \mathcal{D}}
\end{equation}
\begin{equation}
    -(\beta_S + \beta_R)\omega - 2 \beta_S(\alpha N_s + \kappa T) + 2\beta_O(T_s -\alpha \kappa N) |_{\partial \mathcal{D}}= - 2\beta_R |_{\partial\mathcal{D}}
\end{equation}
where
\begin{equation}
    \beta_O = \frac{\eta_O/\mu}{R}.
\end{equation}
Thus, the five dimensionless parameters for the low friction problem are $\alpha$, $\beta_S$, $\beta_R$, $\beta_O$, and $\zeta = H/R$.

The analysis proceeds nearly identically for the high friction Brinkman equation, with one modification: the normal velocity scale $\bar{U}$ is written in terms of the substrate friction, becoming $\gamma/(\Gamma R^2)$. We find the corresponding definitions of $\beta_S, \beta_R$, and $\beta_O$:
\begin{equation}
    \beta_S = \frac{\eta }{\Gamma R^2},\quad\beta_R = \frac{\eta_R }{\Gamma R^2},\quad\beta_O = \frac{\eta_O }{\Gamma R^2}.
\end{equation}
Note that in terms of the penetration depth $\bar\delta$,
\begin{equation}
    \beta_S + \beta_R = \frac{\bar\delta^2}{R^2}.
\end{equation}

\section{An infinite strip}

Here we consider the flow field when $\mathcal{D}$ is an infinite strip of half-width $R$ oriented axially along the $y$-axis. The flow is assumed to be steady and unidirectional. For this section, we will use Cartesian coordinates so that the flow field may be expressed as $\boldsymbol{u}(x,y,z) = v(x,z)\hat{\boldsymbol{y}}$, with $ \boldsymbol{U}(x,y) = \boldsymbol{u}(x,y,0) = V(x)\hat{\boldsymbol{y}}$. This type of flow field satisfies $\bnabla\bcdot \boldsymbol{U} = 0$ on the entire surface, so that Eqn.~(\ref{FTcond}) applies. Defining $g(x) = \mu \partial v/\partial z|_{z=0}$, the $x$-component of the momentum equation inside the monolayer is
\begin{equation}
    -\frac{\mathrm{d}P}{\mathrm{d}x}+ \bar\eta \frac{\mathrm{d}^2V}{\mathrm{d}x^2} = g
\end{equation}
when $|x|<R$. On the other hand, when $|x|>R$, we have $g=0$. At the boundary, $\kappa=0$ and $\hat{\boldsymbol{n}} = \pm \hat{\boldsymbol{x}}$, so the boundary conditions Eqn.~(\ref{stressbcedge}) for this geometry are 
\begin{equation}
    \left. -P\right|_{x= \pm R} = 0 \text{ and } \left. \bar\eta \frac{\mathrm{d}V}{\mathrm{d}x} - 2\eta_R \Omega\right|_{x = \pm R}  =0,
    \label{stripbcs}
\end{equation}
where the outer pressure has been taken to be zero. Analogous to the disc case, the odd viscosity does not enter explicitly in this strip geometry. Note that because the flow is unidirectional, the pressure $P$ inside the domain is harmonic. Since Eqn.~(\ref{stripbcs}) shows $P$ vanishes along its boundary, $P$ must be zero everywhere. The momentum equation is then simply
\begin{equation}
    \bar\eta \frac{\mathrm{d}^2V}{\mathrm{d}x^2}\chi(|x|<R) = g
\end{equation}
Since the flow field has an odd symmetry, we
define $a(k)$ to be the Fourier sine transform of $g(x)$,
\begin{equation}
    a(k)  = \int_{0}^\infty \mathrm{d}x~ g(x) \sin kx \Longleftrightarrow g(x) = \frac{2}{\pi}\int_0^\infty\mathrm{d}k~a(k)\sin kx.
    \label{astrip}
\end{equation}
We take the Fourier sine transform of the momentum equation in $x$ to obtain
\begin{equation}
    \bar\eta\int_{0}^R \mathrm{d}x~ \frac{\mathrm{d}^2V}{\mathrm{d}x^2} \sin kx = a(k).
\end{equation}
Integration by parts yields
\begin{equation}
    2\eta_R\Omega \sin kR - \bar\eta V(R) k \cos kR - \bar\eta k^2 \int_0^R\mathrm{d}x~ V(x)\sin kx = a(k)
    \label{stripie}
\end{equation}
where the fact that $V(x)$ is odd and Eqn.~(\ref{stripbcs}) have been used to simplify boundary terms. Since $\sin z = \sqrt{\pi z/2} J_{1/2}(z)$, this equation is amenable to Tranter's method, which we now demonstrate; the prescription is nearly identical to that of the disc geometry given in \textsection 4.3. Naturally, adapting the Green's function formulation as in \textsection 4.1 yields an identical answer.

\begin{figure}
    \centering
    \includegraphics[scale=0.2]{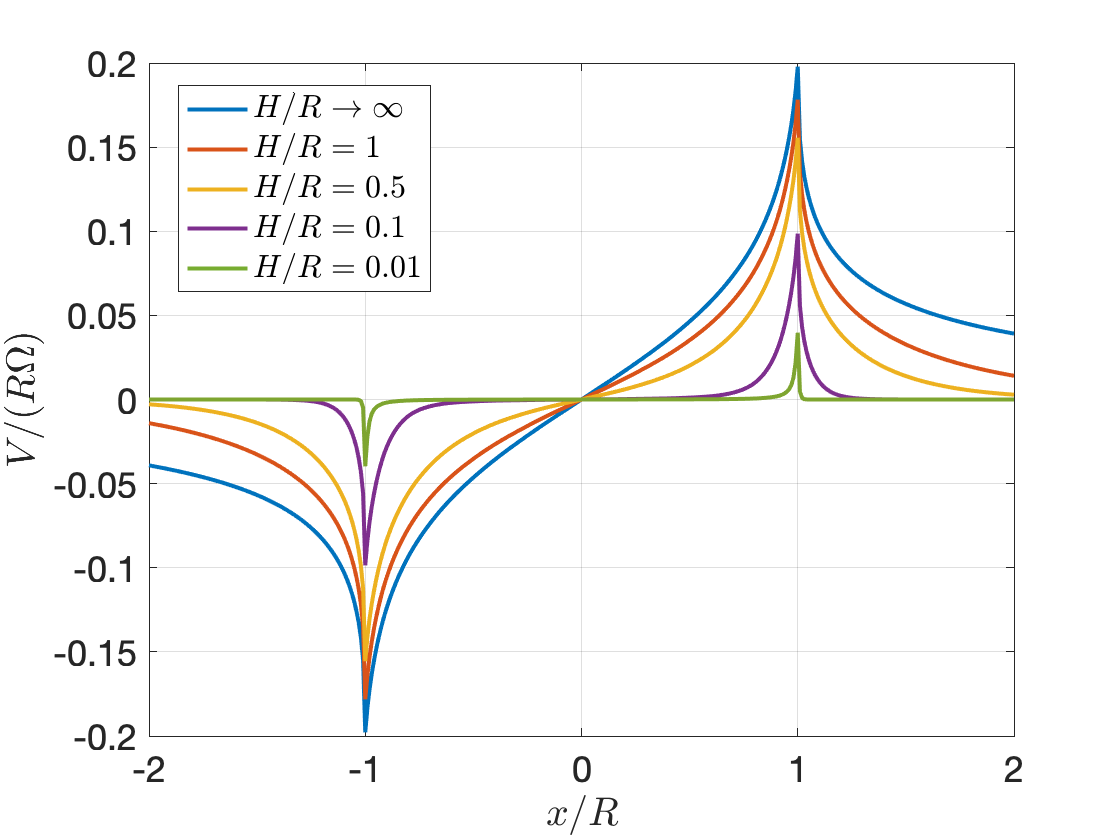}
    \caption{The surface azimuthal velocity $V$ due to an infinite strip of half-width $R$ as subphase depth $H$ is varied. As $H\to 0$, a boundary layer of width $\bar{\delta} = \sqrt{\bar\eta H/\mu}$ becomes visible and $V$ is well-approximated by Eqn.~(\ref{smallHstrip}). Parameters: $\eta_R/\eta =1.875 $, $\mu R/\eta = 30$. Since radial velocity $U=0$ for the disc, the  flow fields are independent of line tension and odd viscosity.}
    \label{fig:strip_vel}
\end{figure}

Adapting~\cite{Tranter1954}, we let
\begin{equation}
    a(k) = k^{1/2-\beta} \sum_{n=0}^\infty a_n J_{2n+1/2+\beta}(kR)
    \label{adefstrip}
\end{equation}
where $\beta>0$ is arbitrary and the coefficients $\{a_n\}$ are unknown. Note that {Eqn.~(\ref{FTcond})} combined with Eqns.~(\ref{astrip}) and (\ref{adefstrip}) implies
\begin{equation}
    V(x) = \frac{2}{\pi\mu} \sum_{n=0}^\infty a_n \int_0^{\infty}  \mathrm{d}k~k^{-1/2-\beta} \tanh kH J_{2n+1/2+\beta}(kR) \sin kx
    \label{Vstrip}
\end{equation}
for the geometry at hand. Substituting this into Eqn.~(\ref{stripie}) gives
\begin{align}
    &2\eta_R\Omega \sin kR-k \cos kR\frac{2\bar\eta}{\pi \mu} \sum_{n=0}^\infty a_n \int_0^\infty \mathrm{d}k'~(k')^{-1/2-\beta} \tanh k'H J_{2n+1/2+\beta}(k'R) \sin k'R \nonumber \\
    &- \frac{2\bar\eta}{\pi \mu} \sum_{n=0}^\infty a_n k^2\int_0^R \mathrm{d}x~ \sin kx \int_0^\infty \mathrm{d}k'~(k')^{-1/2-\beta} \tanh k'H J_{2n+1/2+\beta}(k'R)\sin k'x\nonumber \\
    &= k^{1/2-\beta} \sum_{n=0}^\infty a_n J_{2n+1/2+\beta}(kR)
\end{align}
Multiplying both sides by $k^{-3/2-\beta}J_{2m+1+\beta}(kR)$, where $m$ is a nonnegative integer, integrating from $0$ to $\infty$ in $k$, and interchanging integrals yields the system
\begin{equation}
    2\eta_R\Omega\mu g^{(s)}_m = \sum_{n=0}^\infty a_n\left[\frac{2\bar\eta}{\pi }\xi_m^{(s)} \Lambda_n^{(s)}  + \frac{2\bar\eta}{\pi} M_{nm}^{(s)} + \mu \Delta_{nm}^{(s)}\right],
\end{equation}
where
\begin{equation}
    g^{(s)}_m = \int_0^\infty \mathrm{d}k~k^{-3/2-\beta} J_{2m+1/2+\beta}(kR) \sin kR 
     = \frac{\pi (R/2)^{\beta+1/2}}{2(\beta+1/2)!}\delta_{n0}
\end{equation}
\begin{equation}
    \xi^{(s)}_m = \int_0^\infty \mathrm{d}k~k^{-3/2-\beta} J_{2m+1/2+\beta}(kR) k\cos kR = 0
\end{equation}
\begin{equation}
    \Lambda^{(s)}_n = \int_0^\infty \mathrm{d}k'~(k')^{-1/2-\beta} \tanh k'H J_{2n+1/2+\beta}(k'R)\sin k'R
\end{equation}
\begin{equation}
    M^{(s)}_{mn} = \int_0^\infty \mathrm{d}k'~(k')^{-2\beta} \tanh k'H J_{2n+1/2+\beta}(k'R)J_{2m+1/2+\beta}(k'R)
\end{equation}
\begin{align}
    \Delta^{(s)}_{mn} &= \int_0^\infty \mathrm{d}k~k^{-1-2\beta} J_{2n+1/2+\beta}(kR)J_{2m+1/2+\beta}(kR)\nonumber \\
    &=\frac{\beta(2\beta-1)!(m+n-1/2)!}{4^\beta (\beta+m-n)!(\beta+n-m)!(1/2+2\beta+m+n)!}
\end{align}
and factorials assume their usual definition via the gamma function (Formula 8.310.1 of \cite{GradshteynRyzhik2007}). Note the strong resemblance to Eqns.~(\ref{trantersystem}) to (\ref{deltaeqn}) for the disc. Following the discussion in \textsection 4.3, we choose $\beta = 1/2$ and numerically evaluate $M^{(s)}_{mn}$ and $\Lambda^{(s)}_{n}$ for $m$ and $n< 20$. The truncated linear system is quickly solved for the coefficients $\{a_n\}$. Figure~\ref{fig:strip_vel} depicts the resulting surface flow field $V(x)$ for different values of $H/R$, which is found by evaluating Eqn.~(\ref{Vstrip}).

Finally, we turn to the high friction ($H\to 0$) case, where $V$ satisfies the Brinkman equation
\begin{equation}
    \bar\eta \frac{\mathrm{d}^2 V}{\mathrm{d} x^2}\chi(|x|<R) = \Gamma V,
\end{equation}
with $\Gamma = \mu/H$. Using the boundary condition {Eqn.}~(\ref{stripbcs}), the solution is
\begin{equation}
    V(x) = \frac{ 2\Omega \bar{\delta}\eta_R}{\bar\eta}\frac{\sinh (x/\bar{\delta})}{\cosh (R/\bar{\delta})}\chi(|x|<R),
    \label{smallHstrip}
\end{equation}
where $\bar{\delta} = \sqrt{\bar\eta/\Gamma}$ is the penetration depth of the edge current. The convergence to this solution as $H$ is decreased can be seen in Fig.~\ref{fig:strip_vel}.

\bibliographystyle{jfm}

\bibliography{axisymmetric_droplet}

\begin{thebibliography}{47}
\expandafter\ifx\csname natexlab\endcsname\relax\def\natexlab#1{#1}\fi
\def\au#1{#1} \def\ed#1{#1} \def\yr#1{#1}\def\at#1{#1}\def\jt#1{\textit{#1}}
  \def\bt#1{#1}\def\bvol#1{\textbf{#1}} \def\vol#1{#1} \def\pg#1{#1}
  \def\publ#1{#1}\def\arxiv#1{#1}\def\org#1{#1}\def\st#1{\textit{#1}}

\bibitem[Alexander {\em et~al.\/}(2006)Alexander, Bernoff, Mann, J.~Adin~Mann
  \& Zou]{Alexander_etal2006}
{\sc \au{Alexander, J.~C.}, \au{Bernoff, A.~J.}, \au{Mann, E.~K.},
  \au{J.~Adin~Mann, Jr.} \& \au{Zou, L.}} \yr{2006}  \at{Hole dynamics in
  polymer langmuir films}.  \jt{Phys. Fluids}  \bvol{{\bf 18}},  \pg{062103}.

\bibitem[Alexander {\em et~al.\/}(2007)Alexander, Bernoff, Mann, Mann \&
  Wintersmith]{Alexander_etal2007}
{\sc \au{Alexander, J.~C.}, \au{Bernoff, A.~J.}, \au{Mann, E.~K.}, \au{Mann,
  J.~Adin} \& \au{Wintersmith, J.~R.}} \yr{2007}  \at{Domain relaxation in
  {L}angmuir flims}.  \jt{J. Fluid Mech.}  \bvol{{\bf 571}},  \pg{191--219}.

\bibitem[Avron(1998)]{Avron1998}
{\sc \au{Avron, J.~E.}} \yr{1998}  \at{Odd viscosity}.  \jt{J. Stat. Phys.}
  \bvol{{\bf 92}},  \pg{543}.

\bibitem[Avron {\em et~al.\/}(1995)Avron, Seiler \& Zograf]{Avron_etal1995}
{\sc \au{Avron, J.~E.}, \au{Seiler, R.} \& \au{Zograf, P.~G.}} \yr{1995}
  \at{Viscosity of quantum hall fluids}.  \jt{Phys. Rev. Lett.}  \bvol{75},
  \pg{697--700}.

\bibitem[Berdyugin {\em et~al.\/}(2019)Berdyugin, Xu, Pellegrino,
  Krishna~Kumar, Principi, Torre, Ben~Shalom, Taniguchi, Watanabe, Grigorieva,
  Polini, Geim \& Bandurin]{Berdyugin_etal2019}
{\sc \au{Berdyugin, A.~I.}, \au{Xu, S.~G.}, \au{Pellegrino, F. M.~D.},
  \au{Krishna~Kumar, R.}, \au{Principi, A.}, \au{Torre, I.}, \au{Ben~Shalom,
  M.}, \au{Taniguchi, T.}, \au{Watanabe, K.}, \au{Grigorieva, I.~V.},
  \au{Polini, M.}, \au{Geim, A.~K.} \& \au{Bandurin, D.~A.}} \yr{2019}
  \at{Measuring hall viscosity of graphene{\textquoteright}s electron fluid}.
  \jt{Science}  \bvol{{\bf 364}},  \pg{162}.

\bibitem[Bililign {\em et~al.\/}(2021)Bililign, Balboa~Usabiaga, Ganan, Poncet,
  Soni, Magkiriadou, Shelley, Bartolo \& Irvine]{BililignEtAl2021}
{\sc \au{Bililign, Ephraim~S}, \au{Balboa~Usabiaga, Florencio}, \au{Ganan,
  Yehuda~A}, \au{Poncet, Alexis}, \au{Soni, Vishal}, \au{Magkiriadou, Sofia},
  \au{Shelley, Michael~J}, \au{Bartolo, Denis} \& \au{Irvine, William}}
  \yr{2021}  \at{Motile dislocations knead odd crystals into whorls}.
  \jt{Nature Physics}  \pg{pp. 1--7}.

\bibitem[Busbridge(1938)]{Busbridge1938}
{\sc \au{Busbridge, I.~W.}} \yr{1938}  \at{Dual integral equations}.  \jt{Proc.
  London Math. Soc.}  \bvol{{\bf 44}},  \pg{115--129}.

\bibitem[Cooke(1956)]{Cooke1956}
{\sc \au{Cooke, J.~C.}} \yr{1956}  \at{A solution of {T}ranter's dual integral
  equations problem}.  \jt{Q. J. Mech. Appl. Math.}  \bvol{{\bf 9}},
  \pg{103--110}.

\bibitem[Cooke(1963)]{Cooke1963}
{\sc \au{Cooke, J.~C.}} \yr{1963}  \at{Triple integral equations}.  \jt{Q. J.
  Mech. Appl. Math.}  \bvol{{\bf 16}},  \pg{193--203}.

\bibitem[Cooke(1965)]{Cooke1965}
{\sc \au{Cooke, J.~C.}} \yr{1965}  \at{The solution of triple integral
  equations in operational form}.  \jt{Q. J. Mech. Appl. Math.}  \bvol{{\bf
  18}},  \pg{57--72}.

\bibitem[Cressman {\em et~al.\/}(2004)Cressman, Davoudi, Goldburg \&
  Schumacher]{CG2004}
{\sc \au{Cressman, John~R}, \au{Davoudi, Jahanshah}, \au{Goldburg, Walter~I} \&
  \au{Schumacher, J{\"o}rg}} \yr{2004}  \at{Eulerian and lagrangian studies in
  surface flow turbulence}.  \jt{New Journal of Physics}  \bvol{6}~(1),
  \pg{53}.

\bibitem[Gao {\em et~al.\/}(2017)Gao, Betterton, Jhang \& Shelley]{GBJS2017}
{\sc \au{Gao, Tong}, \au{Betterton, Meredith~D}, \au{Jhang, An-Sheng} \&
  \au{Shelley, Michael~J}} \yr{2017}  \at{Analytical structure, dynamics, and
  coarse graining of a kinetic model of an active fluid}.  \jt{Physical Review
  Fluids}  \bvol{2}~(9),  \pg{093302}.

\bibitem[Gao {\em et~al.\/}(2015)Gao, Blackwell, Glaser, Betterton \&
  Shelley]{GBGBS2015a}
{\sc \au{Gao, Tong}, \au{Blackwell, Robert}, \au{Glaser, Matthew~A},
  \au{Betterton, Meredith~D} \& \au{Shelley, Michael~J}} \yr{2015}
  \at{Multiscale polar theory of microtubule and motor-protein assemblies}.
  \jt{Physical review letters}  \bvol{114}~(4),  \pg{048101}.

\bibitem[Goldburg {\em et~al.\/}(2001)Goldburg, Cressman, V{\"o}r{\"o}s,
  Eckhardt \& Schumacher]{GoldburgEtAl2001}
{\sc \au{Goldburg, WI}, \au{Cressman, JR}, \au{V{\"o}r{\"o}s, Z}, \au{Eckhardt,
  B} \& \au{Schumacher, J}} \yr{2001}  \at{Turbulence in a free surface}.
  \jt{Physical Review E}  \bvol{63}~(6),  \pg{065303}.

\bibitem[Gradshteyn \& Ryzhik(2007)]{GradshteynRyzhik2007}
{\sc \au{Gradshteyn, I.~S.} \& \au{Ryzhik, I.~M.}} \yr{2007} {\em Table of
  Integrals, Series, and Products\/}, 7th edn.  \publ{Burlington, MA:
  Elsevier}.

\bibitem[Held {\em et~al.\/}(1995)Held, Pierrehumbert, Garner \&
  Swanson]{HPGS1995}
{\sc \au{Held, Isaac~M.}, \au{Pierrehumbert, Raymond~T.}, \au{Garner,
  Stephen~T.} \& \au{Swanson, Kyle~L.}} \yr{1995}  \at{Surface
  quasi-geostrophic dynamics}.  \jt{Journal of Fluid Mechanics}  \bvol{282},
  \pg{1–20}.

\bibitem[Henle \& Levine(2009)]{HenleLevine2009}
{\sc \au{Henle, M.~L.} \& \au{Levine, A.~J.}} \yr{2009}  \at{Effective
  viscosity of a dilute suspension of membrane-bound inclusions}.  \jt{Phys.
  Fluids}  \bvol{{\bf 21}},  \pg{033106}.

\bibitem[Jeffery(1915)]{Jeffery1915}
{\sc \au{Jeffery, G.~B.}} \yr{1915}  \at{On the steady rotation of a solid of
  revolution in a viscous fluid}.  \jt{P. Lond. Math. Soc.}  \bvol{{\bf 2}},
  \pg{327--338}.

\bibitem[Jia \& Shelley(2022)]{JiaShelley2022}
{\sc \au{Jia, L.~L.} \& \au{Shelley, M.~J.}} \yr{2022}  \at{The role of
  monolayer viscosity in {L}angmuir film closure dynamics}.  \jt{To be
  submitted} .

\bibitem[Kokot {\em et~al.\/}(2017)Kokot, Das, Winkler, Gompper, Aranson \&
  Snezhko]{Kokot_etal2017}
{\sc \au{Kokot, Gasper}, \au{Das, Shibananda}, \au{Winkler, Roland~G.},
  \au{Gompper, Gerhard}, \au{Aranson, Igor~S.} \& \au{Snezhko, Alexey}}
  \yr{2017}  \at{Active turbulence in a gas of self-assembled spinners}.
  \jt{P. Natl. Acad. Sci. USA}  \bvol{114},  \pg{12870}.

\bibitem[Lubensky \& Goldstein(1996)]{LubenskyGoldstein1996}
{\sc \au{Lubensky, D.~K.} \& \au{Goldstein, R.~E.}} \yr{1996}
  \at{Hydrodynamics of monolayer domains at the air-water interface}.
  \jt{Phys. Fluids}  \bvol{{\bf 8}},  \pg{843}.

\bibitem[Martin \& Smith(2011)]{MartinSmith2011}
{\sc \au{Martin, P.~A.} \& \au{Smith, S. G.~L.}} \yr{2011}  \at{Generation of
  internal gravity waves by an oscillating horizontal disc}.  \jt{Proc. R. Soc.
  A.}  \bvol{{\bf 467}},  \pg{3406--3423}.

\bibitem[Mart{\'\i}nez-Prat {\em et~al.\/}(2019)Mart{\'\i}nez-Prat,
  Ign{\'e}s-Mullol, Casademunt \& Sagu{\'e}s]{MartinezEtAl2019}
{\sc \au{Mart{\'\i}nez-Prat, Berta}, \au{Ign{\'e}s-Mullol, Jordi},
  \au{Casademunt, Jaume} \& \au{Sagu{\'e}s, Francesc}} \yr{2019}  \at{Selection
  mechanism at the onset of active turbulence}.  \jt{Nature physics}
  \bvol{15}~(4),  \pg{362--366}.

\bibitem[Masoud \& Shelley(2014)]{MasoudShelley2014}
{\sc \au{Masoud, H.} \& \au{Shelley, M.~J.}} \yr{2014}  \at{Collective surfing
  of chemically active particles}.  \jt{Phys. Rev. Lett.}  \bvol{{\bf 112}},
  \pg{128304}.

\bibitem[Noble(1958)]{Noble1958}
{\sc \au{Noble, B.}} \yr{1958}  \at{Certain dual integral equations}.  \jt{J.
  Math. Phys.}  \bvol{{\bf 37}},  \pg{128}.

\bibitem[Oppenheimer {\em et~al.\/}(2019)Oppenheimer, Stein \&
  Shelley]{Oppenheimer_etal2019}
{\sc \au{Oppenheimer, Naomi}, \au{Stein, David~B.} \& \au{Shelley, Michael~J.}}
  \yr{2019}  \at{Rotating membrane inclusions crystallize through hydrodynamic
  and steric interactions}.  \jt{Phys. Rev. Lett.}  \bvol{123},  \pg{148101}.

\bibitem[Oppenheimer {\em et~al.\/}(2022)Oppenheimer, Stein, Yah Ben~Zion \&
  Shelley]{Oppenheimer_etal2022}
{\sc \au{Oppenheimer, Naomi}, \au{Stein, David~B.}, \au{Yah Ben~Zion, Matan} \&
  \au{Shelley, Michael~J.}} \yr{2022}  \at{Hyperuniformity and phase enrichment
  in vortex and rotor assemblies}.  \jt{Nature Comm.} .

\bibitem[Petroff {\em et~al.\/}(2015)Petroff, Wu \&
  Libchaber]{Petroff_etal2015}
{\sc \au{Petroff, Alexander~P.}, \au{Wu, Xiao-Lun} \& \au{Libchaber, Albert}}
  \yr{2015}  \at{Fast-moving bacteria self-organize into active two-dimensional
  crystals of rotating cells}.  \jt{Phys. Rev. Lett.}  \bvol{114},
  \pg{158102}.

\bibitem[Pullin(1992)]{Pullin1992}
{\sc \au{Pullin, DI}} \yr{1992}  \at{Contour dynamics methods}.  \jt{Annual
  review of fluid mechanics}  \bvol{24}~(1),  \pg{89--115}.

\bibitem[Ratnanather {\em et~al.\/}(2014)Ratnanather, Kim, Zhang, Davis \&
  Lucas]{Ratnanather_etal2014}
{\sc \au{Ratnanather, J.~T.}, \au{Kim, J.~H.}, \au{Zhang, S.}, \au{Davis, A.
  M.~J.} \& \au{Lucas, S.~K.}} \yr{2014}  \at{Algorithm 935: {IIPBF}, a
  {MATLAB} toolbox for infinite integral of products of two {B}essel
  functions}.  \jt{ACM T. Math. Software}  \bvol{{\bf 40.2}}.

\bibitem[Rodrigo \& Fefferman(2004)]{RF2004}
{\sc \au{Rodrigo, Jos{\'e}~Luis} \& \au{Fefferman, Charles~L}} \yr{2004}
  \at{The vortex patch problem for the surface quasi-geostrophic equation}.
  \jt{Proceedings of the National Academy of Sciences of the United States of
  America}  \pg{pp. 2684--2686}.

\bibitem[Saffman(1995)]{Saffman1995}
{\sc \au{Saffman, Philip~G}} \yr{1995} {\em Vortex dynamics\/}.
  \publ{Cambridge university press}.

\bibitem[Saffman \& Delbr\"uck(1975)]{SaffmanDelbruck1975}
{\sc \au{Saffman, P.~G.} \& \au{Delbr\"uck, M.}} \yr{1975}  \at{Brownian motion
  in biological membranes}.  \jt{P. Natl. Acad. Sci. USA}  \bvol{72}~(8).

\bibitem[Sanchez {\em et~al.\/}(2012)Sanchez, Chen, DeCamp, Heymann \&
  Dogic]{SanchezEtAl2012}
{\sc \au{Sanchez, Tim}, \au{Chen, Daniel~TN}, \au{DeCamp, Stephen~J},
  \au{Heymann, Michael} \& \au{Dogic, Zvonimir}} \yr{2012}  \at{Spontaneous
  motion in hierarchically assembled active matter}.  \jt{Nature}
  \bvol{491}~(7424),  \pg{431--434}.

\bibitem[Sherwood(2013)]{Sherwood2013}
{\sc \au{Sherwood, J.~D.}} \yr{2013}  \at{{S}tokes drag on a disc with a
  {N}avier slip condition near a plane wall}.  \jt{Fluid Dyn. Res.}  \bvol{{\bf
  45}}.

\bibitem[{Sneddon}(1946)]{Sneddon1946}
{\sc \au{{Sneddon}, I.~N.}} \yr{1946}  \at{{The Distribution of Stress in the
  Neighbourhood of a Crack in an Elastic Solid}}.  \jt{Proceedings of the Royal
  Society of London Series A}  \bvol{187}~(1009),  \pg{229--260}.

\bibitem[Sneddon(1966)]{Sneddon1966}
{\sc \au{Sneddon, I.~N.}} \yr{1966} {\em Mixed boundary value problems in
  potential theory\/}, 1st edn.  \publ{Amsterdam: North-Holland Pub. Co.}

\bibitem[Sneddon(1975)]{Sneddon1975}
{\sc \au{Sneddon, I.~N.}} \yr{1975} {\em The use in mathematical physics of
  Erdélyi-Kober operators and of some of their generalizations\/},  \pg{pp.
  37--79}.  \publ{Springer}.

\bibitem[Soni {\em et~al.\/}(2019)Soni, Bililign, Magkiriadou, Sacanna,
  Bartolo, Shelley \& Irvine]{Soni_etal2019}
{\sc \au{Soni, V.}, \au{Bililign, E.}, \au{Magkiriadou, S.}, \au{Sacanna, S.},
  \au{Bartolo, D.}, \au{Shelley, M.~J.} \& \au{Irvine, W. T.~M.}} \yr{2019}
  \at{The free surface of a colloidal chiral fluid: waves and instabilities
  from odd stress and {H}all viscosity}.  \jt{Nat. Phys.}  \bvol{{\bf 15}},
  \pg{1188--1194}.

\bibitem[Souslov {\em et~al.\/}(2019)Souslov, Dasbiswas, Fruchart,
  Vaikuntanathan \& Vitelli]{Souslov_etal2019}
{\sc \au{Souslov, Anton}, \au{Dasbiswas, Kinjal}, \au{Fruchart, Michel},
  \au{Vaikuntanathan, Suriyanarayanan} \& \au{Vitelli, Vincenzo}} \yr{2019}
  \at{Topological waves in fluids with odd viscosity}.  \jt{Phys. Rev. Lett.}
  \bvol{122},  \pg{128001}.

\bibitem[Stone(1995)]{Stone1995}
{\sc \au{Stone, H.~A.}} \yr{1995}  \at{Fluid motion of monomolecular films in a
  channel flow geometry}.  \jt{Phys. Fluids}  \bvol{{\bf 7}},  \pg{2931--2937}.

\bibitem[Stone \& McConnell(1995)]{StoneMcConnell1995}
{\sc \au{Stone, H.~A.} \& \au{McConnell, H.~M.}} \yr{1995}  \at{Hydrodynamics
  of quantized shape transitions of lipid domains}.  \jt{Proc. R. Soc. Lond. A}
   \bvol{{\bf 448}},  \pg{97--111}.

\bibitem[Tranter(1954)]{Tranter1954}
{\sc \au{Tranter, C.~J.}} \yr{1954}  \at{A further note on dual integral
  equations and an application to the diffraction of electromagnetic waves}.
  \jt{Q. J. Mech. Appl. Math.}  \bvol{{\bf 7}},  \pg{317--325}.

\bibitem[Wiegmann \& Abanov(2014)]{WiegmannAbanov2014}
{\sc \au{Wiegmann, Paul} \& \au{Abanov, Alexander~G.}} \yr{2014}  \at{Anomalous
  hydrodynamics of two-dimensional vortex fluids}.  \jt{Phys. Rev. Lett.}
  \bvol{113},  \pg{034501}.

\bibitem[Yan {\em et~al.\/}(2020)Yan, Corona, Malhotra, Veerapaneni \&
  Shelley]{YanEtAl2020}
{\sc \au{Yan, W.}, \au{Corona, E.}, \au{Malhotra, D.}, \au{Veerapaneni, S.} \&
  \au{Shelley, M.~J.}} \yr{2020}  \at{A scalable computational platform for
  particulate {S}tokes suspensions}.  \jt{J. Comput. Physics}  \bvol{416},
  \pg{109524}.

\bibitem[Yan \& Sloan(1988)]{YanSloan1988}
{\sc \au{Yan, Y.} \& \au{Sloan, I.~H.}} \yr{1988}  \at{On integral equations of
  the first kind with logarithmic kernels}.  \jt{J. Integral Equ. Appl.}
  \bvol{1}~(4).

\bibitem[Yeo {\em et~al.\/}(2015)Yeo, Lushi \& Vlahovska]{Yeo_etal2015}
{\sc \au{Yeo, Kyongmin}, \au{Lushi, Enkeleida} \& \au{Vlahovska, Petia~M.}}
  \yr{2015}  \at{Collective dynamics in a binary mixture of hydrodynamically
  coupled microrotors}.  \jt{Phys. Rev. Lett.}  \bvol{114},  \pg{188301}.

\end{thebibliography}


\end{document}